\documentclass[%
 reprint,
superscriptaddress,
 amsmath,amssymb,
 aps,
prd,
longbibliography
]{revtex4-1}

\usepackage{graphicx}
\usepackage{dcolumn}
\usepackage{bm}
\usepackage{hyperref}
\usepackage{color}
\usepackage[dvipsnames]{xcolor}
\usepackage[mathlines]{lineno}
\usepackage{tabularx}
\usepackage{subfigure}
\usepackage[normalem]{ulem}
\usepackage[export]{adjustbox}
\usepackage{url}
\usepackage{listings}
\hypersetup{
    pdfnewwindow=true,    
    colorlinks=true,      
    linkcolor=blue,       
    citecolor=blue,       
    filecolor=blue,       
    urlcolor=blue         
}

\definecolor{green}{rgb}{0, 0.6, 0}

\newcolumntype{Y}{>{\centering\arraybackslash}X}

\usepackage{natbib}
\begin{document}
\title{Decaying dark matter in dwarf spheroidal galaxies: \\ Prospects for X-ray and gamma-ray telescopes}


\author{Shin'ichiro Ando}
\email{s.ando@uva.nl}
\thanks{\scriptsize \!\! \href{http://orcid.org/0000-0001-6231-7693}{orcid.org/0000-0001-6231-7693}}
\affiliation{GRAPPA Institute, University of Amsterdam, 1098 XH Amsterdam, The Netherlands}
\affiliation{Institute for Theoretical Physics, University of Amsterdam, 1098 XH Amsterdam, The Netherlands}
\affiliation{Kavli Institute for the Physics and Mathematics of the Universe (Kavli IPMU, WPI), University of Tokyo, Kashiwa, Chiba 277-8583, Japan}

\author{Suvendu K. Barik}
\affiliation{Institute for Theoretical Physics, University of Amsterdam, 1098 XH Amsterdam, The Netherlands}

\author{Zhuoran Feng}
\affiliation{GRAPPA Institute, University of Amsterdam, 1098 XH Amsterdam, The Netherlands}

\author{Marco Finetti}
\affiliation{Institute for Theoretical Physics, University of Amsterdam, 1098 XH Amsterdam, The Netherlands}

\author{Andreas Guerra Chaves}
\affiliation{GRAPPA Institute, University of Amsterdam, 1098 XH Amsterdam, The Netherlands}

\author{Sahaja Kanuri}
\affiliation{GRAPPA Institute, University of Amsterdam, 1098 XH Amsterdam, The Netherlands}

\author{Jorinde Kleverlaan}
\affiliation{Anton Pannekoek Institute for Astronomy, University of Amsterdam, 1098 XH Amsterdam, The Netherlands}

\author{Yixuan Ma}
\affiliation{Anton Pannekoek Institute for Astronomy, University of Amsterdam, 1098 XH Amsterdam, The Netherlands}

\author{Nicolo Maresca Di Serracapriola}
\affiliation{Institute for Theoretical Physics, University of Amsterdam, 1098 XH Amsterdam, The Netherlands}

\author{Matthew S. P. Meinema}
\affiliation{GRAPPA Institute, University of Amsterdam, 1098 XH Amsterdam, The Netherlands}

\author{Imanol Navarro Martinez}
\affiliation{Institute for Theoretical Physics, University of Amsterdam, 1098 XH Amsterdam, The Netherlands}

\author{Kenny C. Y. Ng}
\email{kcyng@cuhk.edu.hk}
\thanks{\scriptsize \!\! \href{http://orcid.org/0000-0001-8016-2170}{orcid.org/0000-0001-8016-2170}}
\affiliation{GRAPPA Institute, University of Amsterdam, 1098 XH Amsterdam, The Netherlands}
\affiliation{Institute for Theoretical Physics, University of Amsterdam, 1098 XH Amsterdam, The Netherlands}
\affiliation{Department of Physics, The Chinese University of Hong Kong, Shatin, Hong Kong China}

\author{Ebo Peerbooms}
\affiliation{Institute for Theoretical Physics, University of Amsterdam, 1098 XH Amsterdam, The Netherlands}

\author{Casper A. van Veen}
\affiliation{GRAPPA Institute, University of Amsterdam, 1098 XH Amsterdam, The Netherlands}

\author{Fabian Zimmer}
\affiliation{GRAPPA Institute, University of Amsterdam, 1098 XH Amsterdam, The Netherlands}

\date{March 24, 2021}

\begin{abstract}
Dwarf spheroidal galaxies are dark matter dominated systems, and as such, ideal for indirect dark matter searches.
If dark matter decays into high-energy photons in the dwarf galaxies, they will be a good target for current and future generations of X-ray and gamma-ray telescopes.
By adopting the latest estimates of density profiles of dwarf galaxies in the Milky Way, we revise the estimates dark matter decay rates in dwarf galaxies; our results are more robust, but weaker than previous estimates.  Applying these results, we study the  detectability of dark matter decays with X-ray and very-high-energy gamma-ray telescopes, such as eROSITA, XRISM, Athena, HAWC, and CTA.
Our projection shows that all of these X-ray telescopes will be able to critically assess the claim of the 7~keV sterile neutrino decays from stacked galaxy clusters and nearby galaxies. For TeV decaying dark matter, we can constrain its lifetime to be longer than $\sim$10$^{27}$--10$^{28}$~s.  We also make projections for future dwarf galaxies that would be newly discovered with the Vera Rubin Observatory Legacy Survey of Space and Time, which will further improve the expected sensitivity to dark matter decays both in the keV and PeV mass ranges.
\end{abstract}


\maketitle

\section{Introduction}
\label{sec:Introduction}


Although the existence of dark matter is firmly established through its gravitational interactions, its particle nature with non-gravitational interactions is yet to be unveiled.
Intensive effort to search for weakly interactive massive particles (WIMPs) --- a prime candidate for particle dark matter for a long time --- has unfortunately yielded null detection so far from all the avenues of search strategies: colliders, direct and indirect experiments~(\cite{Roszkowski:2017nbc}, see also Ref.~\cite{Leane:2018kjk}).
This has significant increased interests in other dark matter candidates in recent years~\cite{Boyarsky:2018tvu, Graham:2015ouw, Tulin:2017ara, Ishiwata:2019aet}.
Furthermore, although it is often assumed that dark matter is stable as in the case of WIMPs, it does not have to be completely stable as long as its lifetime is much longer than the age of the Universe.
In this paper, we discuss two such possibilities: keV sterile neutrinos (e.g., \cite{Boyarsky:2018tvu}; Sec.~\ref{sub:Sterile neutrino dark matter}) and heavy dark matter with masses in the TeV--PeV range (e.g., \cite{Murase:2015gea, Ishiwata:2019aet}; Sec.~\ref{sub:Heavy dark matter}).
As a target source to look for signatures of dark matter decay, we consider dwarf spheroidal galaxies (dSphs) --- satellites of the Milky-Way halo (Sec.~\ref{sub:Dwarf spheroidal galaxies}).

\subsection{Sterile neutrino dark matter}
\label{sub:Sterile neutrino dark matter}

Sterile neutrinos, $\nu_s$, whose mass is in keV range is a popular example of decaying dark matter~\cite{Boyarsky:2018tvu}.
If they are dark matter, they exist as a mass eigenstate with mostly right-handed (hence sterile) component with a tiny mixture of active flavors such as $\nu_e$.
The sterile neutrinos then decay by emitting a photon ($\nu_s \to \nu_e + \gamma$), which carry energy corresponding to half of the sterile neutrino mass, $E_\gamma = m_{\nu_s}/2$.

There has been a claim of a possible detection of an X-ray line that potentially originated from the sterile neutrino decay.
Reference~\cite{Bulbul:2014sua} looked at the stacked X-ray spectra of 73 galaxy clusters and found an excess emission around 3.5~keV. By splitting up the sample, they ruled out the possibility of it coming solely from nearby bright clusters. The emission line was interpreted as a signature of sterile neutrino decay with mass of $m_{\nu_s} = 7.1$~keV and mixing angle of $\sin^2(2\theta) = 7 \times 10^{-11}$.  A similar claim was made by Ref.~\cite{Boyarsky:2014jta} by looking for the signal from the M~31 and Perseus cluster.

This claim triggered multiple follow-up papers, intense debates, re-investigating the excess at the 3.5 keV, or also trying to explain it with different phenomena (e.g., \cite{Jeltema:2014qfa, Bulbul:2014ala, Jeltema:2014mla, Malyshev:2014xqa, Tamura:2014mta, Aharonian:2016gzq,Perez:2016tcq, Cappelluti:2017ywp, Ng:2019gch, Roach:2019ctw}).
A recent study~\cite{Gall:2019vib} found some missing features in the originally used database near the unidentified line. The inclusion of these features raises the total flux value in the region around 3.5~keV but is still not enough to explain the line found in the stacked galaxy cluster spectra.
There is also an ongoing debate triggered by the null detection with joint-likelihood analysis and stringent constraints claimed by Ref.~\cite{Dessert:2018qih} and commented by Refs.~\cite{Abazajian:2020unr, Boyarsky:2020hqb}.
All of these show that it is still very much an open debate and no concrete conclusions can be drawn yet, without tackling this problem from every possible angle (also see Ref.~\cite{Foster:2021ngm} for the most recent stringent limits).

One way to further investigate this open problem is to use upcoming X-ray telescopes such as eROSITA~\cite{Merloni:2012uf}, XRISM~\cite{XRISMScienceTeam:2020rvx}, and Athena~\cite{Barret:2018qft}. In this paper, we study sensitivities of these telescopes to the sterile neutrino decay. XRISM and Athena have superb energy resolution that is ideal for line searches, while eROSITA enables the all-sky survey. They will provide a deeper look needed to potentially end this debate. 

\subsection{Heavy dark matter}
\label{sub:Heavy dark matter}

The dark matter can be much heavier than GeV--TeV --- typical mass scales of the WIMP dark matter.
This makes an interesting possibility to be tested in light of current and upcoming data of TeV gamma-ray~(HAWC~\cite{Albert:2017vtb}, CTA~\cite{Knodlseder:2020onx}, and LHAASO~\cite{Bai:2019khm}) and and TeV-PeV neutrino telescopes~(IceCube~\cite{Murase:2015gea}, IceCube-Gen2~\cite{Dekker:2019gpe}, KM3NeT~\cite{Dekker:2019gpe, Ng:2020ghe}). 

For dark matter in a mass range of 1~TeV--10~PeV, the most stringent constraints on the decay lifetime of these particles was obtained by the Fermi Large Area Telescope, which are in the range of $10^{28}$--$10^{29}$~s (see Ref.~\cite{Ishiwata:2019aet} and references therein).

These limits can be further improved with current and next generation TeV gamma-ray telescopes. HAWC features a wide field of view, and detects both gamma-rays and cosmic rays in an energy range between $\sim$500~GeV and a few hundred TeV~\cite{Abeysekara:2017mjj}. It consists of 300 water Cherenkov tanks that detect particles from air showers.
CTA is a next-generation gamma-ray and cosmic ray observatory. Its field of view is up to $10^{\circ}$. It consists of two parts, indicated as CTA North and CTA South, which are complementary to each other for the sky coverage. It will cover a broad energy rage from about 20~GeV to 300~TeV~\cite{Knodlseder:2020onx}.
The Large High Altitude Air Shower Observatory (LHAASO)~\cite{Bai:2019khm} is a near-future wide field of view gamma-ray observatory that is nearly completed~\cite{Aharonian:2020iou, LHAASO:2021zta}, which will be sensible to gamma rays in the range between 300~GeV and 1~PeV.  Compared with other gamma-ray observatories, LHAASO will also have improved sensitivities above 50~TeV. In this work, we focus on the prospects with HAWC and CTA.

\subsection{Dwarf spheroidal galaxies}
\label{sub:Dwarf spheroidal galaxies}

dSphs provide a good environment to test the particle nature of dark matter.
Due to its proximity and relatively dense environments, they can be one of the best places to indirectly look for dark matter non-gravitational interactions, such as decay into X-ray or gamma-ray photons.
In addition, because of the paucity of stars, gas, and any other baryonic components, if any signals were detected, they would hardly be coming from astrophysical components, hence making them an ideal target for dark matter searches.

In this paper, we make predictions for constraints on dark matter decay using current and future generation of telescopes of both X-rays and gamma rays, focusing on sterile neutrinos and heavy dark matter, respectively.
We will study implications by using both the currently known dSphs and those that would be discovered by future surveys such as Vera C. Rubin Observatory Legacy Survey of Space and Time (LSST) (\url{www.lsst.org}).
The LSST will cover the entire southern sky and find dozens of new dSphs~\cite{Drlica-Wagner:2019xan, Ando:2019rvr}.

The rate of dark matter decay depends on the density profiles of dSphs.
By applying theories of structure formation to the evolution of subhalos and satellites, Ref.~\cite{Ando:2020yyk} revised the estimate of density profiles of known dwarf galaxies.
The effect was found substantial in terms of rates of WIMP annihilation, by lowering the previous estimates on upper limits to the annihilation cross section by a factor of 2--7.
The same consideration will also reduce the decay rates.
By using the same subhalo and satellite models, we are able to predict dSphs that would be discovered with the LSST, and also their density profiles~\cite{Ando:2019rvr}.
Assessing these quantitatively and also giving the most precise estimates for the current and future indirect dark matter search strategies of various wavebands with all the dSphs are the goals of this work.

The rest of the paper is organized as follows.
In Sec.~\ref{sec:Dark matter decays in dwarf galaxies}, we briefly introduce essential formulae in order to calculate the X-ray and gamma-ray flux from dark matter decay in dSphs.
Section~\ref{sec:Distribution of dwarf D factor} discusses distributions of decay rates in known dSphs (Sec.~\ref{sub:Dark matter decay in known dwarfs}) and LSST dSphs (Sec.~\ref{sub:Dark matter decay in LSST dwarfs}), investigating dependence on various parameters related to dSph formation in dark matter subhalos.
Our main results about the sensitivities of X-ray and gamma-ray telescopes are discussed in Secs.~\ref{sec:sterile_neutr_DM} and \ref{sec:Heavy dark matter} for sterile neutrinos and heavy dark matter, respectively.
We then conclude the paper in Sec.~\ref{sec:Conclusions}.

\section{Dark matter decays in dwarf galaxies}
\label{sec:Dark matter decays in dwarf galaxies}

The differential flux of photons from dark matter decay from a sky region with a solid angle $\Delta \Omega$ is given by
\begin{equation}
    \frac{dF}{dE} = \frac{\Gamma_\chi}{4\pi m_\chi}\frac{dN_{\rm decay}}{dE}D,
    \label{eq:dFdE}
\end{equation}
where $m_\chi$ and $\Gamma_\chi$ are the mass and decay width of dark matter particle $\chi$, respectively, and $dN_{\rm decay}/dE$ is the energy spectrum of the particle of interest per decay.
The decay lifetime $\tau_\chi$ is related to $\Gamma_\chi$ via $\tau_\chi = \Gamma_\chi^{-1}$.
The flux is proportional to the so-called astrophysical $D$ factor:
\begin{equation}
    D_{\rm dSph} = \int_{\Delta\Omega} d\Omega \int d\ell \rho_\chi(r(\ell,\psi)),
    \label{eq:D factor}
\end{equation}
where $d\Omega = 2\pi \sin\psi d\psi$ and $\psi$ is an angle coordinate subtending from the center of the dSph.
We assume that the dark matter density is approximated by a spherically-symmetric Navarro-Frenk-White (NFW) profile~\cite{Navarro:1996gj}:
\begin{equation}
    \rho_\chi(r) = \frac{\rho_s}{(r/r_s)(r/r_s+1)^2},
    \label{eq:NFW}
\end{equation}
up to a tidal truncation radius $r_t$, beyond which $\rho_{\chi} = 0$.
Therefore, density profiles for each dwarf galaxy are characterized by three parameters: $r_s$, $\rho_s$, and $r_t$.
For each set of $(r_s, \rho_s, r_t)$ and the integration angle $\alpha_{\rm int}$, we compute the $D$ factor using a fitting formula found by Ref.~\cite{Evans:2016xwx} as follows:
\begin{eqnarray}
    D_{\rm dSph} &=& \frac{4\pi\rho_s r_s^3}{d^2}
 \left[\ln\left(\frac{{\rm min}[\alpha_{\rm int}d,r_t]}{2r_s}\right)    \right.\nonumber\\&&{}\left.+X\left(\frac{{\rm min}[\alpha_{\rm int}d,r_t]}{r_s}\right)\right],
 \label{eq:D factor fit}
\end{eqnarray}
where $d$ is a distance to the dSph, and 
\begin{equation}
    X(s) = \left\{\begin{array}{ll}
    {\rm arcsech}\,{s}/\sqrt{1-s^2}, & 0\le s \le 1,\\
    {\rm arcsec}\,{s}/\sqrt{s^2-1}, & s \ge 1.\\
    \end{array}\right.
\end{equation}

We also note that the $D$ factor in Eq.~(\ref{eq:dFdE}) has another contribution from the smooth Galactic halo component.
It can be computed by the same formulae, Eqs.~(\ref{eq:D factor}) and (\ref{eq:NFW}), but with specific input parameters like $r_s = 20$~kpc, $r_{t}$ set to be the virial radius $R_\odot = 200$~kpc, and $\rho_s$ chosen to yield the local dark mater density of 0.4~GeV~cm$^{-3}$ at Galactocentric radius of the solar system, $r = R_\odot = 8.5$~kpc~\cite{Pato:2015dua}.
In order to obtain the $D$ factor from the Milky-Way halo component, $D_{\rm MW}$, we first calculate the line-of-sight integral of $\rho_\chi$ at a specific position in the sky of central coordinate of the dSph and then multiply it by the solid angle $\Delta \Omega$.
For this, we evaluate the Galactocentric radius $r$ as a function of the line-of-sight coordinate $\ell$ and the dSph's sky location $\psi$ as 
\begin{equation}
r = \sqrt{R_\odot^2+\ell^2-2R_\odot \ell\cos\psi}.
\end{equation}
The total $D$ factor for the dwarf is then obtained as
$D = D_{\rm dSph} + D_{\rm MW}$.

\section{Distribution of dwarf \textit{D} factor}
\label{sec:Distribution of dwarf D factor}

\subsection{Dark matter decay in known dwarfs}
\label{sub:Dark matter decay in known dwarfs}

The density profile of the dwarf galaxies are obtained through stellar kinematics observations (e.g., \cite{2019ARA&A..57..375S}).
However, especially for ultrafaint dwarf galaxies, where there are not enough data, prior distributions of density profile parameters such as $r_s$ and $\rho_s$ have to be adopted in order to obtain their meaningful constraints in the Bayesian parameter inference.
Often in the literature, non-informative priors --- i.e., uniform distributions for both $\ln r_s$ and $\ln \rho_s$ --- were used (e.g.,~\cite{Geringer-Sameth:2014yza}).

Instead, in this paper, we follow the procedure in Ref.~\cite{Ando:2019rvr} to estimate the density profile of observed dwarf galaxies, and also the expected distribution of dwarf $D$ factors.
The procedure employs subhalo models that account for their accretion onto the Milky-Way host halo at given redshift and mass, followed by tidal stripping after accretion. The model employed here was developed semi-analytically in Ref.~\cite{Hiroshima:2018kfv}, which was calibrated to the results of the N-body numerical simulations~\cite{Ishiyama:2014gla,Makiya:2015spa}. 
We adopt the analysis results of Ref.~\cite{Ando:2020yyk} to obtain the posterior distribution of $(r_s, \rho_s, r_t)$ for each known dwarf galaxy.
Then the distribution of $D$ factors for each dwarf is computed by using Eq.~(\ref{eq:D factor fit}). However, there still remains a source of uncertainty within this prior modelling approach, which is connected to the condition that a satellite galaxy forms in a subhalo.
We parameterize this with the peak value of the maximum circular velocity in the subhalo at the satellite formation, $V_{\rm peak}$.
We adopt $V_{\rm peak}>14~\rm km~s^{-1}$ as the canonical condition, but investigate a range of different values and their impact on the results.
For classical dSphs such as Draco, we instead adopt $V_{\rm peak}>25~\rm km~s^{-1}$ prior~\cite{Ando:2020yyk}.
Lastly, we compute the Milky-Way halo contribution, $D_{\rm MW}$, based on the sky locations of each of the known dwarfs.

\begin{figure}
    \centering
    \includegraphics[width=8.5cm]{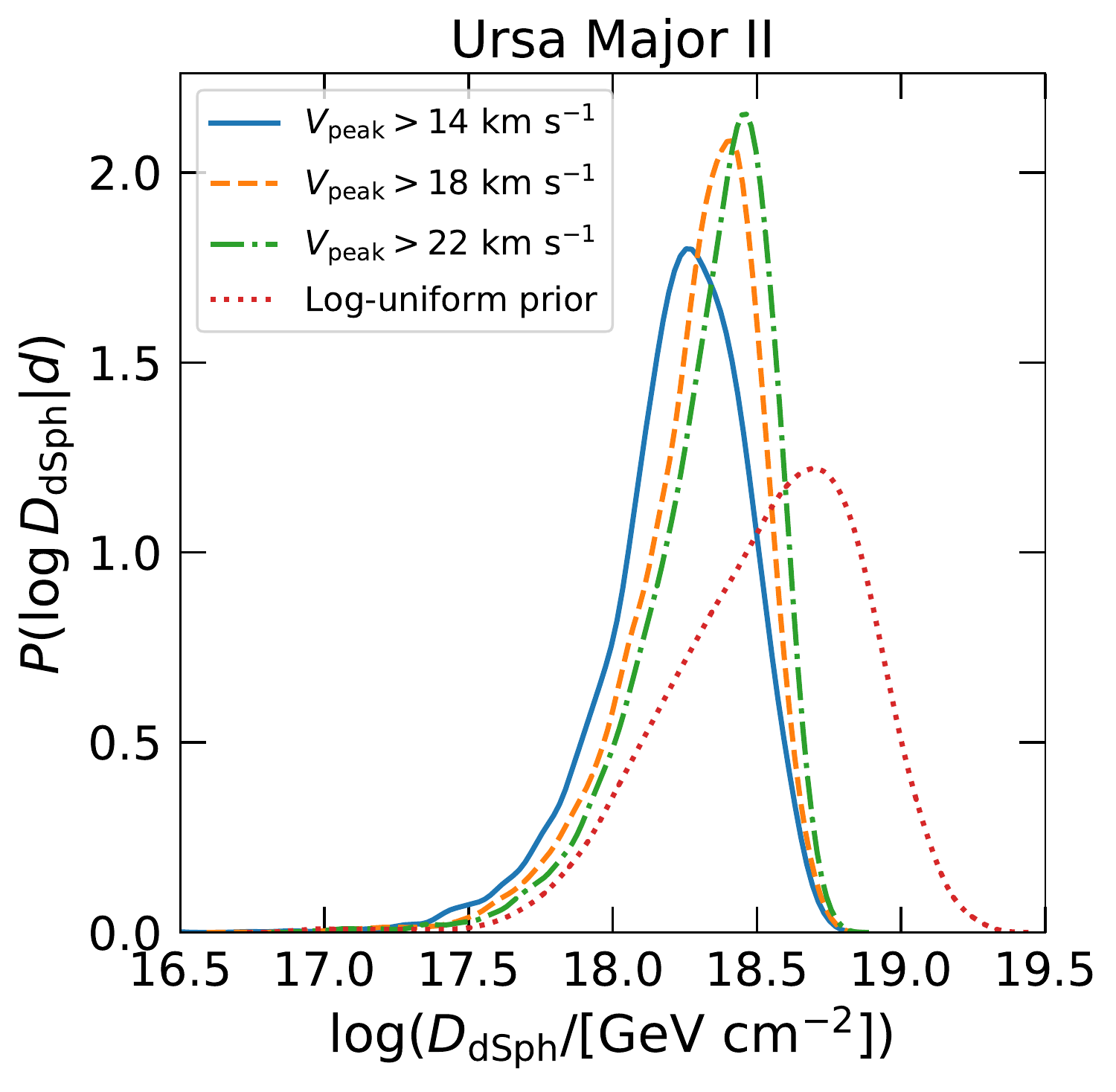}
    \caption{Posterior probability distributions of $D_{\rm dSph}(0.5^{\circ})$ for Ursa Major II, obtained with satellite priors with $V_{\rm peak}>14~\rm km~s^{-1}$ (solid), $V_{\rm peak}>18~\rm km~s^{-1}$ (dashed), $V_{\rm peak}>22~\rm km~s^{-1}$ (dot-dashed), and log-uniform priors (dotted).}
    \label{fig:D_distribution_UMa_II}
\end{figure}

The resulting $D$-factor distributions for the different prior assumptions studied here are presented in Fig.~\ref{fig:D_distribution_UMa_II} for the ultrafaint dSph, Ursa Major II. We find that physically motivated priors yield narrower posterior distributions with smaller median than the uninformative prior, as found in Ref.~\cite{Ando:2020yyk}.
Median and 68\% credible regions for $\log(D_{\rm dSphs}/[\rm GeV~cm^{-2}])$ integrated out to 0.5$^{\circ}$ with these posteriors are $18.25_{-0.26}^{+0.20}$ ($V_{\rm peak}>14$~km~s$^{-1}$), $18.32_{-0.25}^{+0.17}$ ($V_{\rm peak}>18$~km~s$^{-1}$), $18.38_{-0.28}^{+0.16}$ ($V_{\rm peak}>22$~km~s$^{-1}$), and $18.58_{-0.38}^{+0.30}$ (log-uniform prior).
We also note that, for the Ursa Major II, the Milky-Way halo contribution gives $\sim$60\% of the total $D$ factor.
If $D_{\rm dSph}$ is obtained by integrating up to 0.05$^{\circ}$ instead, the corresponding median and 68\% credible regions for $\alpha_{\rm int}=0.05^{\circ}$ are $16.75_{-0.17}^{+0.15}$ ($V_{\rm peak}>14$~km~s$^{-1}$), $16.80_{-0.15}^{+0.13}$ ($V_{\rm peak}>18$~km~s$^{-1}$), $16.83_{-0.14}^{+0.13}$ ($V_{\rm peak}>22$~km~s$^{-1}$), and $17.00_{-0.19}^{+0.17}$ (log-uniform prior).
In this case, given that the NFW profile features central cusp, the contribution from the dSph is more important, with the fractional contribution from the Milky-Way halo decreases to about $\sim$30\%.

\begin{figure*}
    \centering
    \includegraphics[width=17cm]{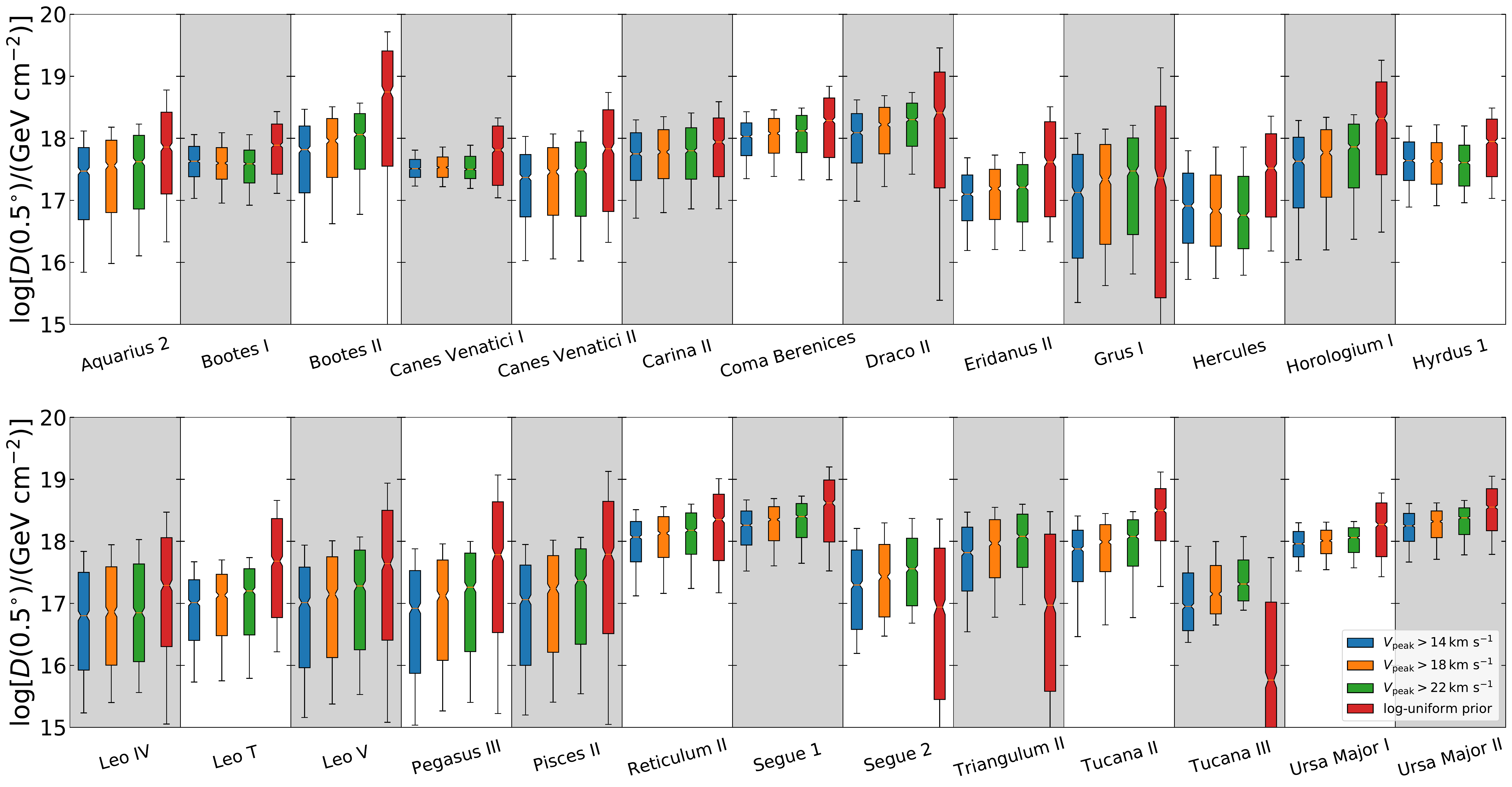}
    \caption{Box-whisker diagram displaying the median and 68\% (boxes) and 95\% (whiskers) credible intervals of the $D$-factor posterior distribution for ultrafaint dSphs, with the integration angle up to $0.5^{\circ}$.  The intervals corresponding to the four satellite priors considered in this research are presented as $V_{\rm peak}>14~\rm km~s^{-1}$ (blue), $V_{\rm peak}>18~\rm km~s^{-1}$ (orange), $V_{\rm peak}>22~\rm km~s^{-1}$ (green), and log-uniform (red).}
    \label{fig:groupA_boxplot}
\end{figure*}

The medians and the 68\% and 95\% credible intervals of these same priors for all the known ultrafaint dwarf galaxies are displayed in Fig.~\ref{fig:groupA_boxplot}. 
There we observe overall similar features as for Fig.~\ref{fig:D_distribution_UMa_II}, with sharper predictions from informative priors and a moderate trend of smaller $D$-factor estimates compared with uninformative prior.

\begin{figure}
    \centering
    \includegraphics[width=8.5cm]{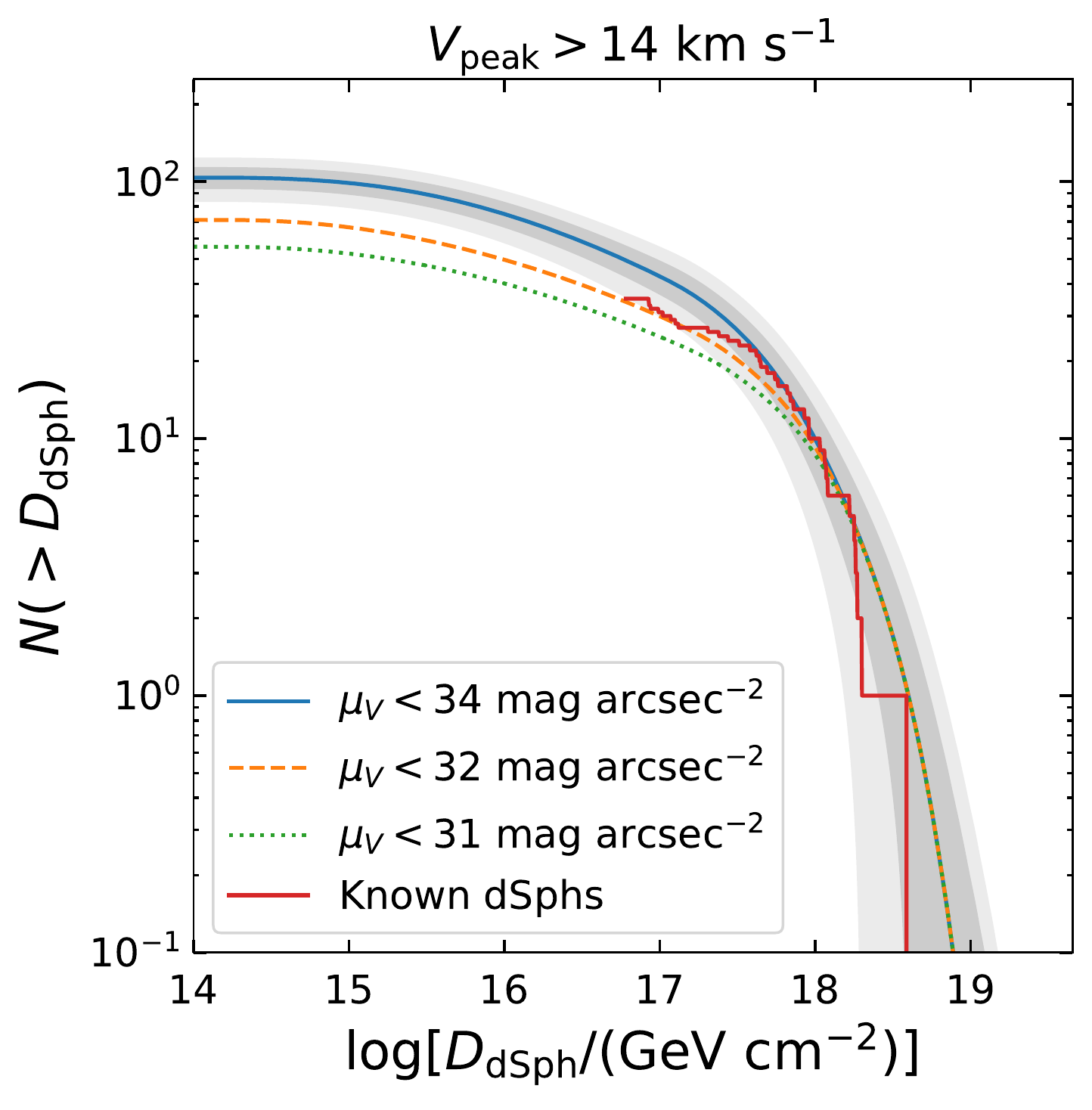}
    \caption{Cumulative number of all-sky satellite dSphs with $V_{\rm peak} > 14$~km~s$^{-1}$ for a few values of $\mu_{V}$ threshold. The grey shaded region represents the Poisson errors at $1\sigma$ and $2\sigma$ levels. The distribution of median $D_{\rm dSph}$ values of the known dSphs is also shown for comparison.}
    \label{fig:groupB_variable_MuV}
\end{figure}

\begin{figure}
    \centering
    \includegraphics[width=8.5cm]{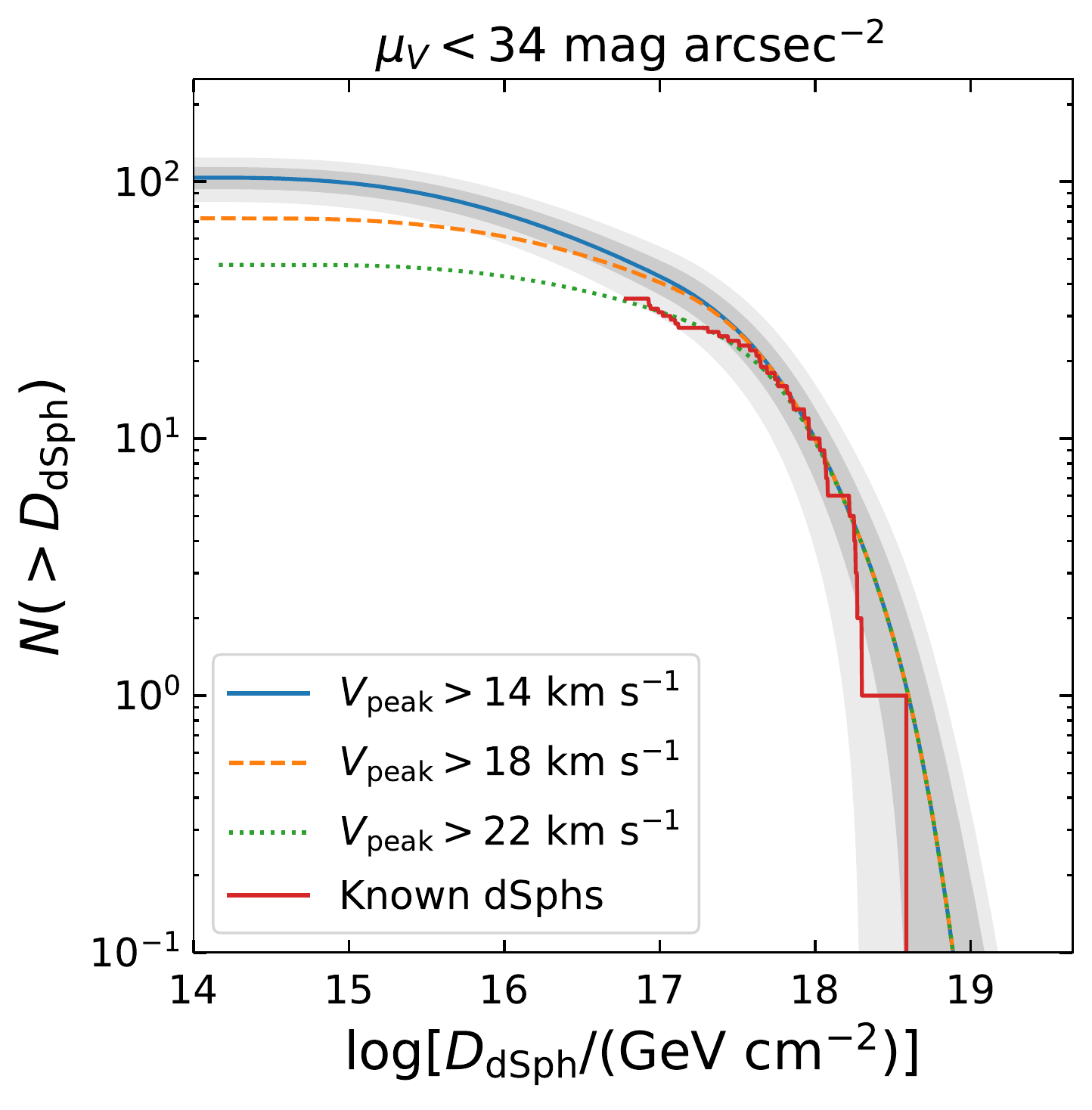}
    \caption{The same as Fig.~\ref{fig:groupB_variable_MuV} but for $\mu_{V} < 34$~mag~arcsec$^{-2}$ and various values of $V_{\rm peak}$ threshold.}
    \label{fig:groupB_variable_Vpeak}
\end{figure}

The medians and both 68\% and 95\% credible intervals for all the ultrafaint dSphs studied in this work is summarized in Fig.~\ref{fig:groupA_boxplot}. 
One can see a general trend that the satellite priors yield systematically smaller $D_{\rm dSph}$, whereas the effect is not as prominent as in the case of the annihilation studied in Ref.~\cite{Ando:2020yyk}.
The results of $D_{\rm dSph}$ distributions of all the dSphs are also discussed in Appendix~\ref{app:Summary of dark matter decay in known dwarf galaxies} in greater details.

\subsection{Dark matter decay in LSST dwarfs}
\label{sub:Dark matter decay in LSST dwarfs}

LSST will discover many more ultrafaint dwarf galaxies from the entire southern sky. 
Many of the known dSphs are found in the northern sky, and the LSST will therefore be able to complement this by searching for many more faint dSphs in the southern sky that is largely unexplored yet. Bright and nearby dwarfs in the southern sky may help further constrain models for indirect dark matter signals~\cite{Ando:2019rvr}.
In order to obtain a $D$-factor distribution of these potential dSphs and then to estimate detectability of signals from decaying dark matter, it is necessary to obtain both a spatial distribution and a distribution of expected subhalo properties.
While Ref.~\cite{Ando:2019rvr} explored the dark matter annihilation (and the so-called $J$ factor), the procedure for modeling the subhalos and their satellite galaxies is similar to the one described in the previous subsection.
As for the spatial distribution, we adopt the results of the hydrodynamical simulations of Ref.~\cite{Calore:2016ogv}, combined with the correction for baryonic disruption and completeness effects from Ref.~\cite{Kim:2017iwr}.

However, several assumptions have to be made in order to connect the subhalos and their satellite galaxies with given luminosity. 
We implement a cutoff on the $V$-band surface brightness that corresponds to the LSST sensitivity, and also $V_{\rm peak}$ corresponding to the formation threshold for dSphs. As in the previous subsection, the canonical value for $V_{\rm peak}$ threshold is taken to be 14~km~s$^{-1}$, below which dSphs are assumed not to form.

In Figs.~\ref{fig:groupB_variable_MuV} and \ref{fig:groupB_variable_Vpeak}, we show cumulative distribution of $D_{\rm dSph}$ from the {\it all sky}, $N(>D_{\rm dSph})$ by varying thresholds for the $V$-band surface brightness ($\mu_V$) and $V_{\rm peak}$, respectively. 
The surface brightness condition of $\mu_V < 32$~mag~arcsec$^{-2}$ corresponds to the expectation for the LSST Year 1 data.
We show Poisson uncertainties of these distributions (both 1$\sigma$ and 2$\sigma$ levels) as grey bands, with uncertainties calculated as $\sigma_{N} = \sqrt{N(>D)}$. 
We also included a cumulative distribution of all known dSph $D$ factors (median values obtained in the previous subsection for $V_{\rm peak} > 14$~km~s$^{-1}$), overlayed in these two figures. We note that the cumulative distribution of the known dSphs follows closely our predictions for the LSST dSphs, in good agreement within the Poisson errors at large-$D$ regime.

For $\mu_V<34$~mag~arcsec$^{-2}$ and $V_{\rm peak}>14$~km~s$^{-1}$, the mean number of dSphs in the all sky is $\sim$100. We first generate the number of subhalos that satisfy this condition through Monte Carlo simulation following the Poisson distribution with this mean. Then, following the $D_{\rm dSph}$ distribution, we assign $D_{\rm dSph}$ values to each of these mock dSphs. We also generate the sky location for each of these dSphs based on the column density of subhalos along each direction. If they are in the survey footprint of the LSST (i.e., declination smaller than 5 degrees), the dSph will be discovered with the LSST.

\begin{figure}
    \centering
    \includegraphics[width=8.5cm]{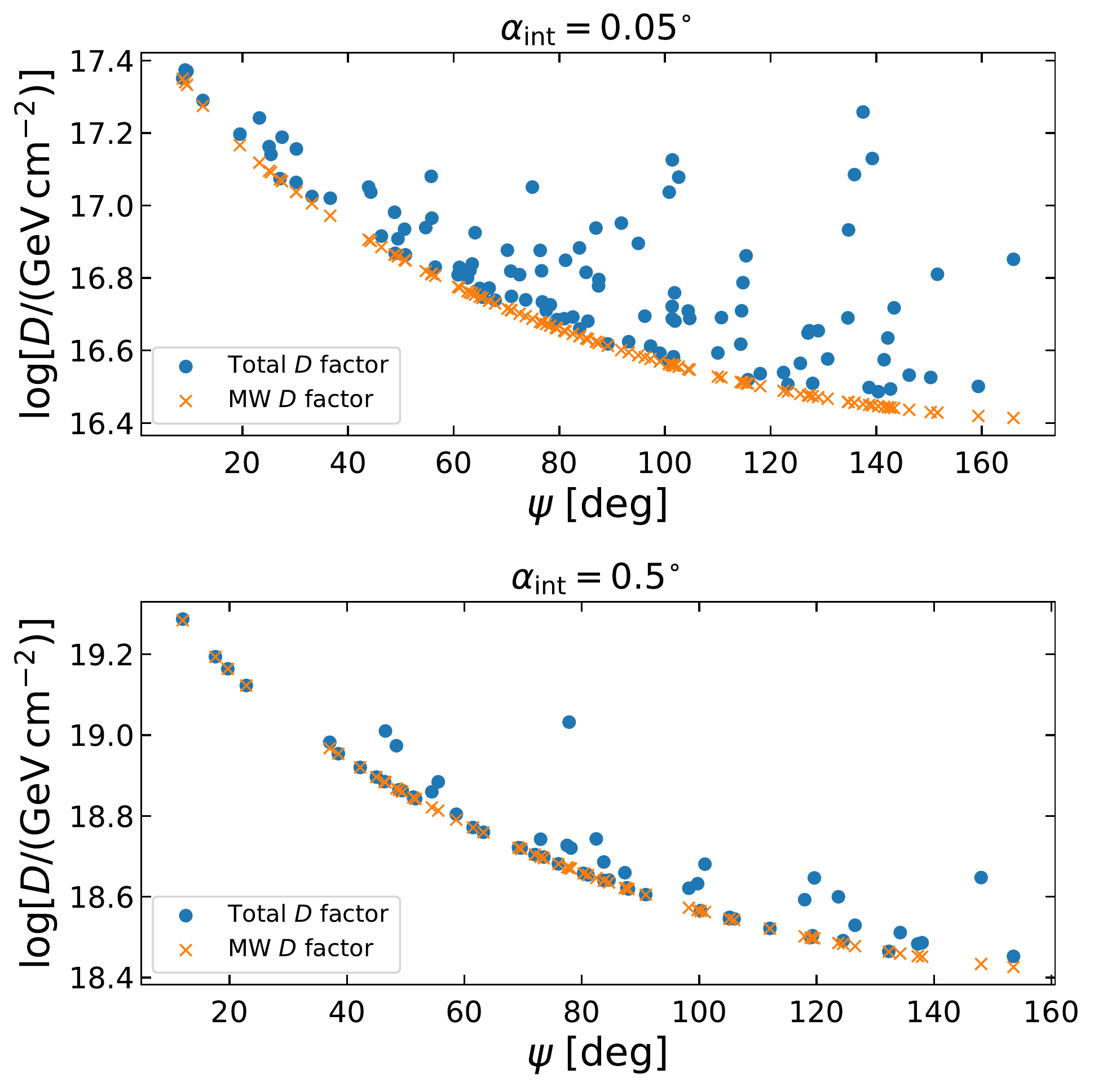}
    \caption{$D$ factors generated by Monte Carlo simulations for mock LSST dSphs and contribution of the Milky-Way halo component as a function of angle $\psi$ from the Galactic center. Top and bottom panels show two cases of different integration angle: $\alpha_{\rm int}$ = 0.05$^\circ$ and 0.5$^{\circ}$, respectively. Orange crosses show the contribution of the Milky-Way halo, whereas the blue dots show the total $D$ factor, i.e. the sum of the dSph and Milky-Way components.}
    \label{fig:DFactor_alpha_int_dependence}
\end{figure}

In addition to this, based on the sky location of each LSST dSph, we compute the Milky Way component of the $D$ factor, $D_{\rm MW}$. The relative contribution of this component to the total $D$ factor depends strongly on the integration angle $\alpha_{\rm int}$. In Fig.~\ref{fig:DFactor_alpha_int_dependence}, we show composition of the $D$ factors for two values of the integration angle. For $\alpha_{\rm int}=0.5^\circ$ a large fraction of the dSph $D$ factors are nearly indistinguishable from the Milky-Way component, while for $\alpha_{\rm int} = 0.05^\circ$, we see that most dSph $D$ factors are strongly separated from the Milky-Way component. 

\section{Sterile neutrino dark matter}
\label{sec:sterile_neutr_DM}

Sterile neutrinos communicate with the standard model only via mixing with active left-handed neutrino species. The conversion probability of a sterile neutrino into an active neutrino is proportional to $\sin^2 2\theta$, where $\theta$ is the mixing angle. The relation between the sterile neutrino decay width $\Gamma_{\nu_s} (= \tau_{\nu_s}^{-1})$ and mixing angle is (e.g.,~\cite{Boyarsky:2018tvu})
\begin{equation}
   \Gamma_{\nu_s}(m_{\nu_s}, \theta) = 1.38 \times 10^{-29}\ \text{s}^{-1} \left( \frac{\sin^2 2 \theta}{10^{-7}} \right) \left( \frac{m_{\nu_s}}{1\ \text{keV}} \right)^5 . 
   \label{eq:Gamma}
\end{equation}

The X-ray flux is given by Eq.~(\ref{eq:dFdE}), and as the X-ray photons are produced through the decay process $\nu_s \to \nu_e + \gamma$, the energy spectrum per decay ${dN_{\rm decay}}/{dE}$ is a delta function:
\begin{equation}
    \frac{dN_{\rm decay}}{dE} = \delta\left(E-\frac{m_{\nu_s}}{2}\right).
\end{equation}
With this flux, we can calculate the event counts in an energy bin between $E_1$ and $E_2$ with
\begin{equation}  \label{eq:event counts (N)}
    N = T \int_{E_1}^{E_2} dE A_{\rm eff}(E) \int dE' P(E, E') \frac{dF}{dE'} ,
\end{equation}
where $P(E,E')$ takes the detector energy resolution into account as a probability of assigning an energy $E$ to an event with true energy $E'$, $A_{\rm eff}(E)$ is the effective area of the detector, and $T$ is its exposure time.
For the energy resolution, we adopt a normal distribution:
\begin{equation}
    P(E,E') = \frac{1}{\sqrt{2\pi} \sigma_E} \exp \left[-\frac{\left({E-E'} \right)^2}{2\sigma_E^2} \right],
\end{equation}
which is characterized by detector specific value of $\sigma_E$. For Athena, XRISM, and eROSITA, they have an energy resolution in terms of the full width at the half maximum ($\text{FWHM} \equiv 2\sqrt{2 \ln 2} \sigma_E$) of 2.5~eV~\cite{Barret:2018qft}, 7~eV~\cite{XRISMScienceTeam:2020rvx} and 138~eV~\cite{Merloni:2012uf}, respectively.
We summarize this and all the other relevant specifications for each of these detectors in Table.~\ref{tab:groupC_table_telescope_params}.

\setlength{\tabcolsep}{10pt}
\begin{table*}
    \scriptsize
	\caption{Summary of specifications for Athena~\cite{Barret:2018qft}, XRISM~\cite{XRISMScienceTeam:2020rvx} and eROSITA~\cite{Merloni:2012uf}.}
	\begin{tabular}{ l c c c c c c }
		\hline
		\hline
		& Athena & XRISM & eROSITA \\
		\hline
		Energy resolution (FWHM) & 2.5 eV & 7 eV & 138 eV \\
		Effective area at 3.5 keV & 4367 $\mathrm{cm^2}$ & 219 $\mathrm{cm^2}$ & 554 $\mathrm{cm^2}$ \\
		Detector background & $5.8 \times 10^3 \, \mathrm{keV^{-1}\,s^{-1}\,sr^{-1}}$ & $2 \times 10^3 \, \mathrm{keV^{-1}\,s^{-1}\,FOV^{-1}}$ & $1151\,\mathrm{keV^{-1}\,s^{-1}\,sr^{-1}}$ \\
		Angular resolution & 2.5 arcmin & 1.7 arcmin (HPD) & $< 15''$ on axis (HEW @ 1.5\,keV) \\
		Field of view (FOV) & 5 arcmin (diameter) & $2.9 \times 2.9$ arcmin$^{2}$ & 0.833 deg$^{2}$ \\
		\hline
		\hline
	\end{tabular}
\label{tab:groupC_table_telescope_params}
\end{table*}

To the detector background in Table~\ref{tab:groupC_table_telescope_params}  (which are energy-independent in given units), we added a contribution coming from the cosmic X-ray background~\cite{Lumb:2002sw}. The contribution depends on X-ray photon energy, which becomes significant at lower energies compared with the detector background.

In order to assess sensitivity to sterile neutrino decays, for each detector, we generate mock data set ($n_i$; the subscript $i$ runs over different energy bins) through Monte Carlo simulations with Poisson distribution assuming null hypothesis, where there is no dark matter component (i.e., $\Gamma = 0$).
The theoretically predicted signal and background counts in each bin $i$, as a function of decay width $\Gamma$ is then marked as $\mu_i(\Gamma)$. The likelihood, i.e., the probability of obtaining the mock data $n_i$ given the parameter $\Gamma$, is then calculated as a product of Poisson probability mass functions with the mean $\mu_i(\Gamma)$, $P\left[n_i|\mu_i(\Gamma)\right]$:
\begin{equation}
    \mathcal{L}(\Gamma) = \prod_i P\left[n_i|\mu_i(\Gamma)\right] = \prod_i \frac{\mu_i(\Gamma)^{n_i}e^{-\mu_i(\Gamma_{\nu_s})}}{n_i!}.
    \label{eq:likelihood}
\end{equation}
The test statistic (TS) used in the prediction of the decay rate is defined as:
\begin{equation}
    \text{TS}(\Gamma) = -2\ln \left[ \frac{\mathcal{L}(\Gamma)}{\mathcal{L}_{\text{max}}}\right],
    \label{eq:TS}
\end{equation}
where $\mathcal{L}_{\text{max}}$ refers to the maximum likelihood, and the corresponding decay width would be the best fit parameter $\Gamma_0$. An acceptable hypothesis of decay width with 95\% confidence level (CL) requires $\text{TS}\leq 2.71$, which determines that the upper limit of the decay rate appears when $\text{TS} = 2.71$.

\subsection{Results for pointing instruments: XRISM and Athena}

\begin{figure*}[htbp!]
	\centering
	\includegraphics[width=8.5cm]{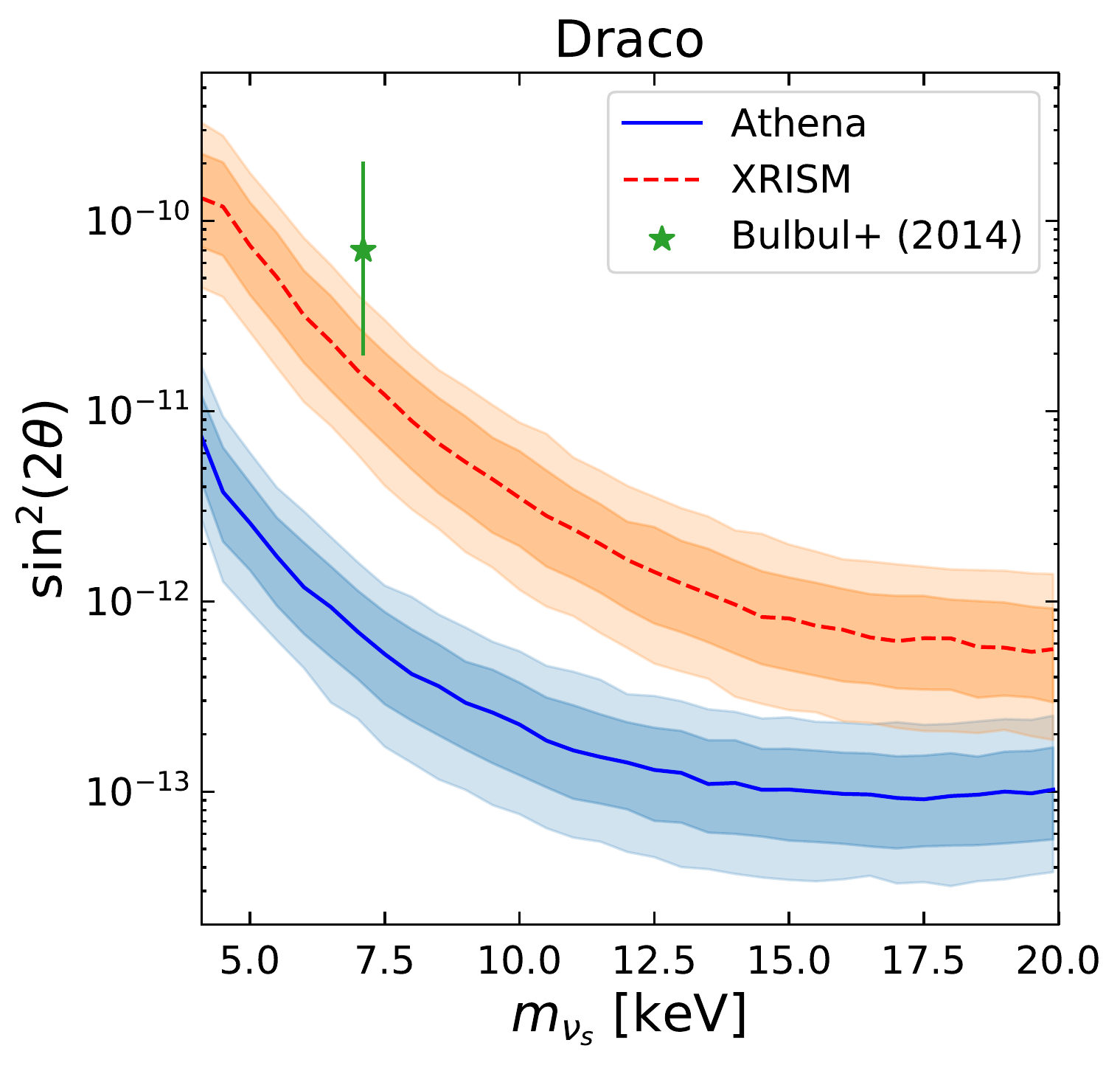}
	\includegraphics[width=8.5cm]{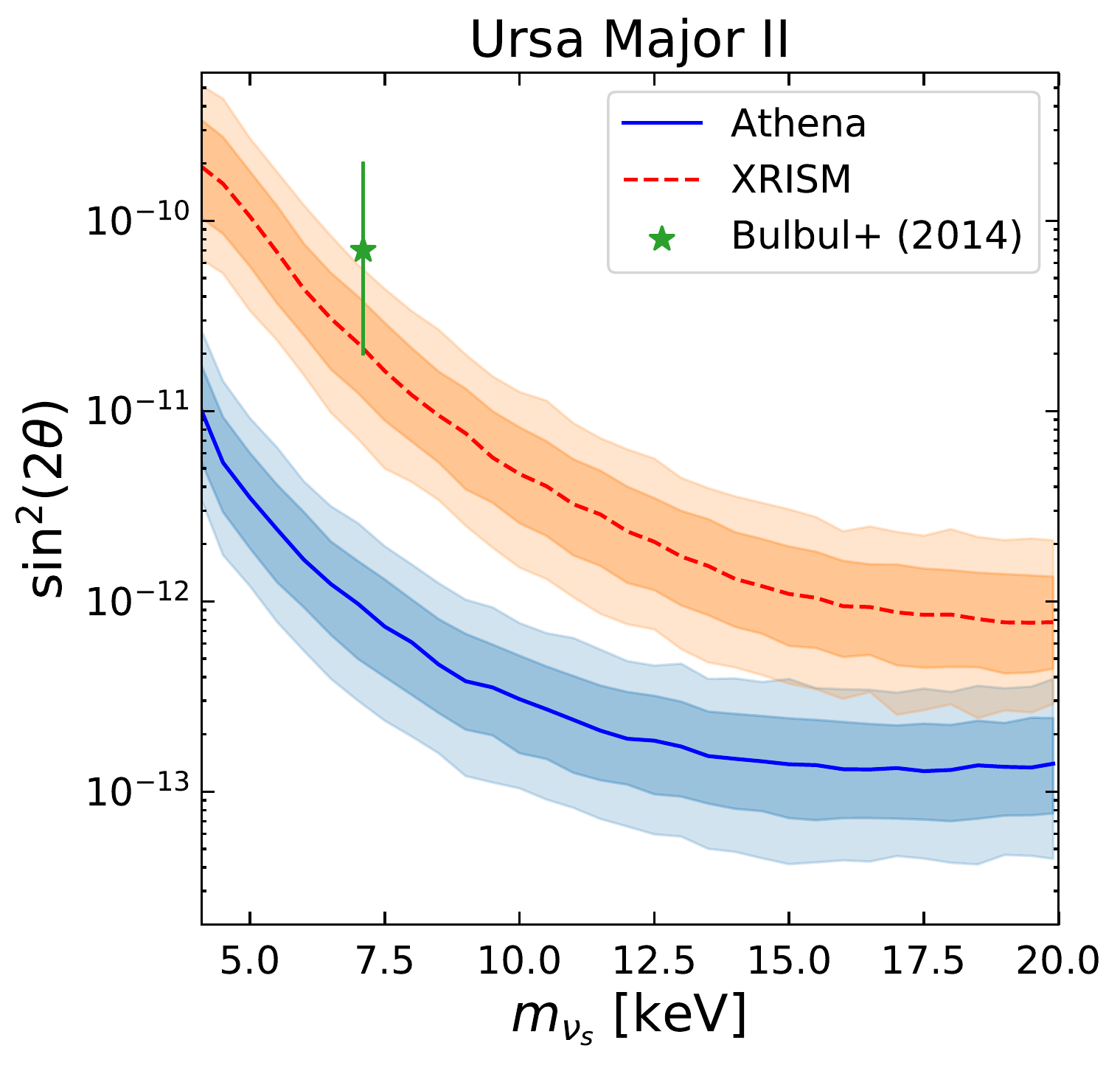}
	\caption{Upper limits (95\% CL) on mixing angle $\theta$ as a function of sterile neutrino mass $m_{\nu_s}$ for XRISM and Athena using $D$-factor distribution from Draco (left) and Ursa Major II (right). The thick and thin bands show 68\% and 95\% containment regions, respectively, accommodating uncertainty of $D$ factor for each dSph as well as the Poisson fluctuation of the X-ray photon counts. In both panels, the half opening angle for Athena is $2.5\, \text{arcmin}$, while for XRISM, $1.64\, \text{arcmin}$. The best-fit parameter region for claimed possible 3.5~keV line~\cite{Bulbul:2014sua} is also shown for comparison.}
	\label{fig:groupC_Athena_XRISM}
\end{figure*}

In Sec.~\ref{sub:Dark matter decay in known dwarfs}, we discussed the distribution of $D$ factors for each known dSph, based on which a Monte-Carlo sample is generated. Randomly choosing a value from the sample of $D$ factors, we calculate a corresponding 95\% CL upper limit on $\Gamma_{\nu_s}$.
We repeat this procedure and obtain distribution of the $\Gamma_{\nu_s}$ upper limit. With Eq.~(\ref{eq:Gamma}), the results are then converted to the limits on $\sin^2 2\theta$.
Figure~\ref{fig:groupC_Athena_XRISM} shows the 95\% CL upper limits on $\sin^2 2\theta$ as a function of sterile neutrino masses $m_{\nu_s}$ both for XRISM and Athena, which would be obtained by observing two dSphs, Draco (left) and Ursa Major II (right). The thick and thin bands show 68\% and 95\% containment regions, respectively, around the medians shown with curves in the middle. The uncertainties come from both those of $D$ factor estimates and the Poisson fluctuation of the X-ray photon count.
We also show the best-fit parameter region of claimed 3.5-keV X-ray line~\cite{Bulbul:2014sua}, and this study clearly shows that the claim can be definitely assessed with these future instruments by looking at promising dSphs. 

\begin{figure}
	\centering
	\includegraphics[width=8.5cm]{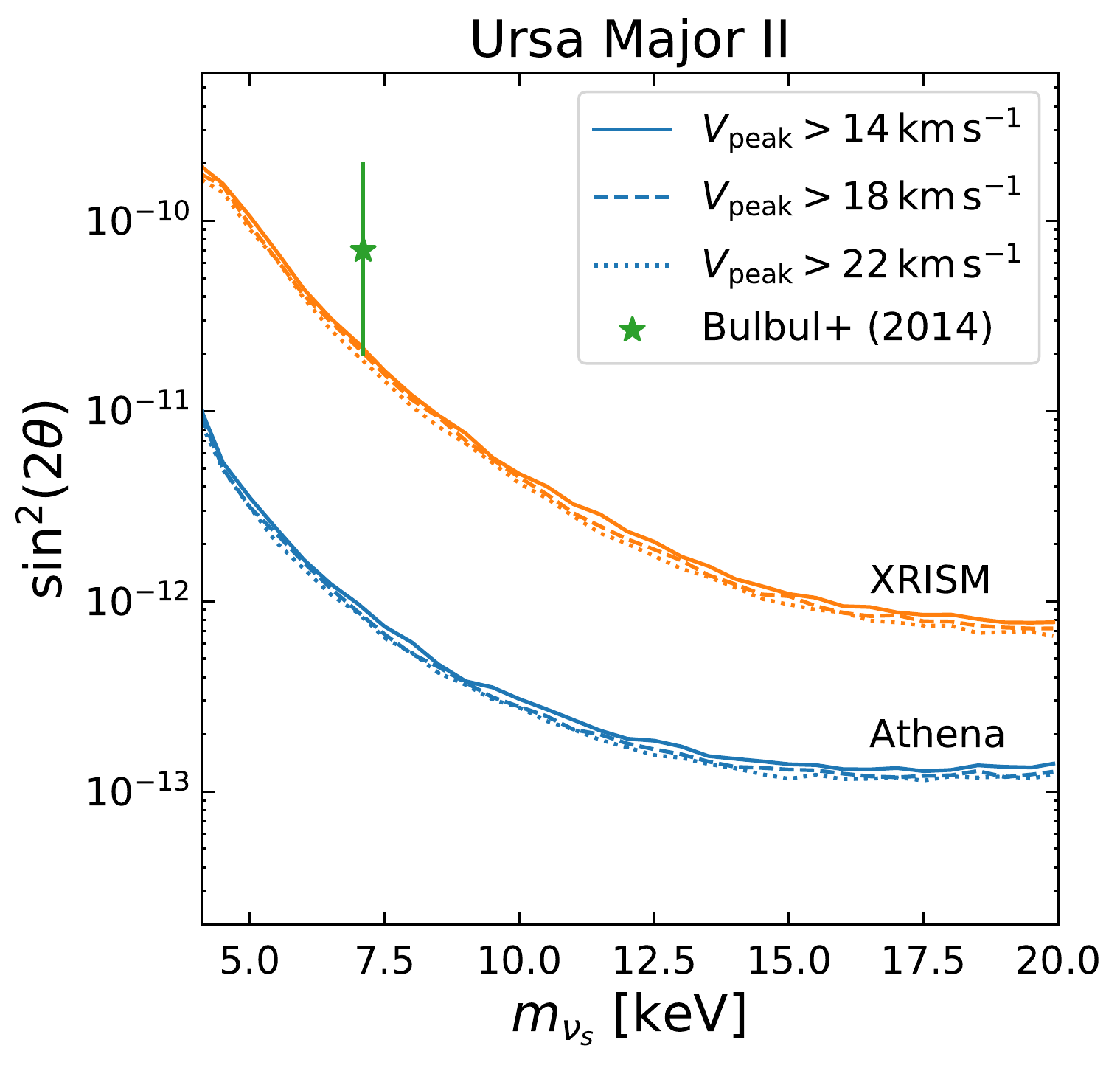}
	\caption{Mixing angle constraints for Athena and XRISM with Ursa Major II ultrafaint dSph, where only medians are shown for different satellite formation conditions parameterized with $V_{\rm peak}$.}
	\label{fig:groupC_Vpeak}
\end{figure}

In this work, following Ref.~\cite{Ando:2020yyk}, in order to obtain the $D$-factor distribution of the ultrafaint dSphs, we modeled subhalos based on cold dark matter framework. Sterile neutrinos, however, are considered instead to behave as warm dark matter, where small-scale structures tend to be erased. 
This will impact the constraints on the mixing angle through prior distribution of density profile and hence of $D$ factors.
However, satellite-formation condition, which we parameterized as $V_{\rm peak}$ thresholds is little known.
As they both change the number of small and faint satellites, the effect of warm dark matter power spectrum is entirely degenerate with the $V_{\rm peak}$ threshold.
To this end, we effectively test the impact of warm dark matter power spectrum by adopting larger values of $V_{\rm peak}$ threshold such as 18 and 22~km~s$^{-1}$, shown in Fig.~\ref{fig:groupC_Vpeak} for the ultrafaint dSph Ursa Major II.
We find very little dependence on different satellite forming conditions through different values of $V_{\rm peak}$ threshold, well within the uncertainty bands shown in Fig.~\ref{fig:groupC_Athena_XRISM}.
This justifies that the model of cold dark matter subhalos combined with phenomenological treatment of satellite formation can be used to discuss sterile neutrino constraints with dSphs.

\subsection{Results for all-sky instrument: eROSITA}

\begin{figure*}[htbp!]
    \centering
    \includegraphics[width=8.5cm]{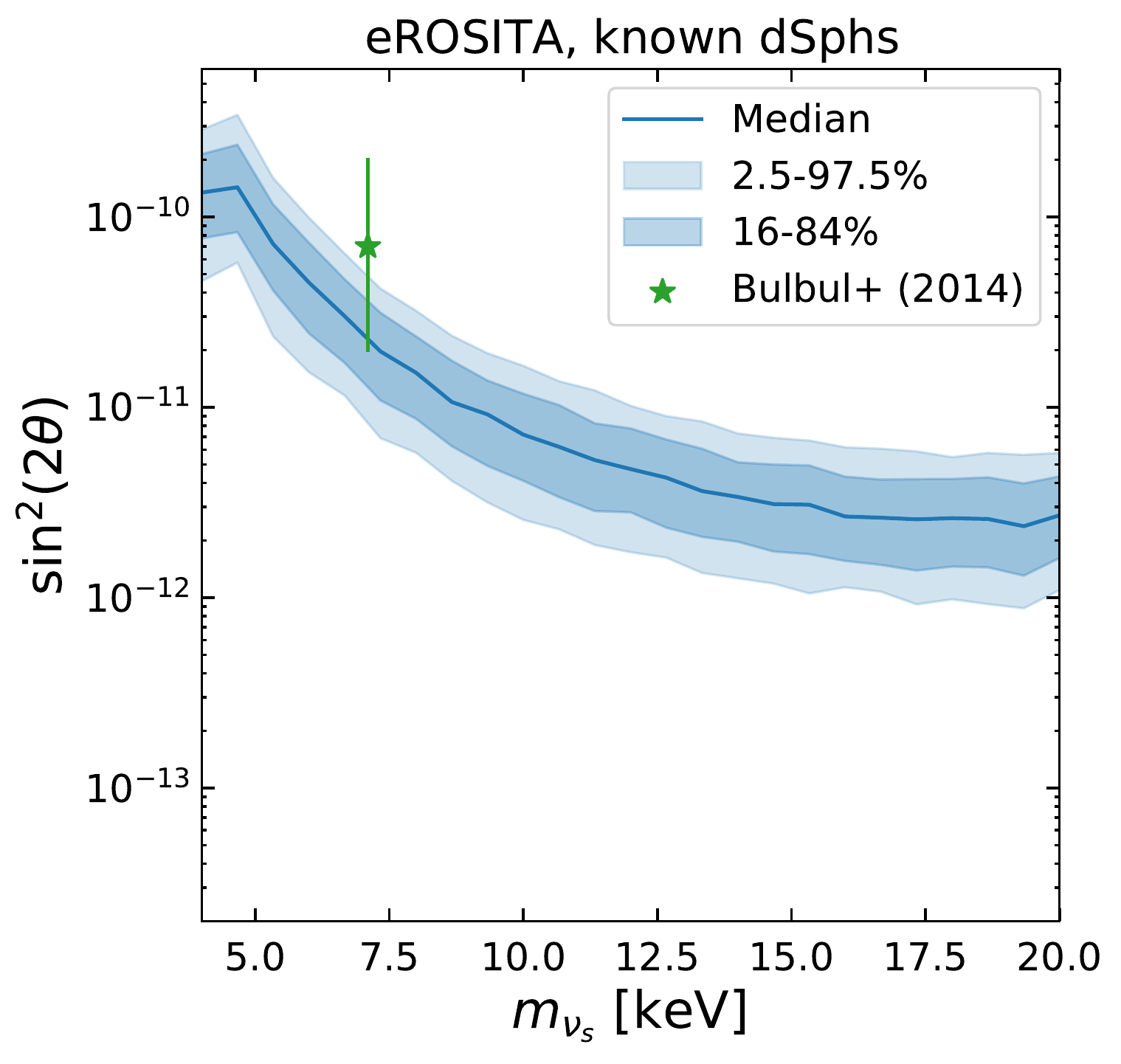}
    \includegraphics[width=8.5cm]{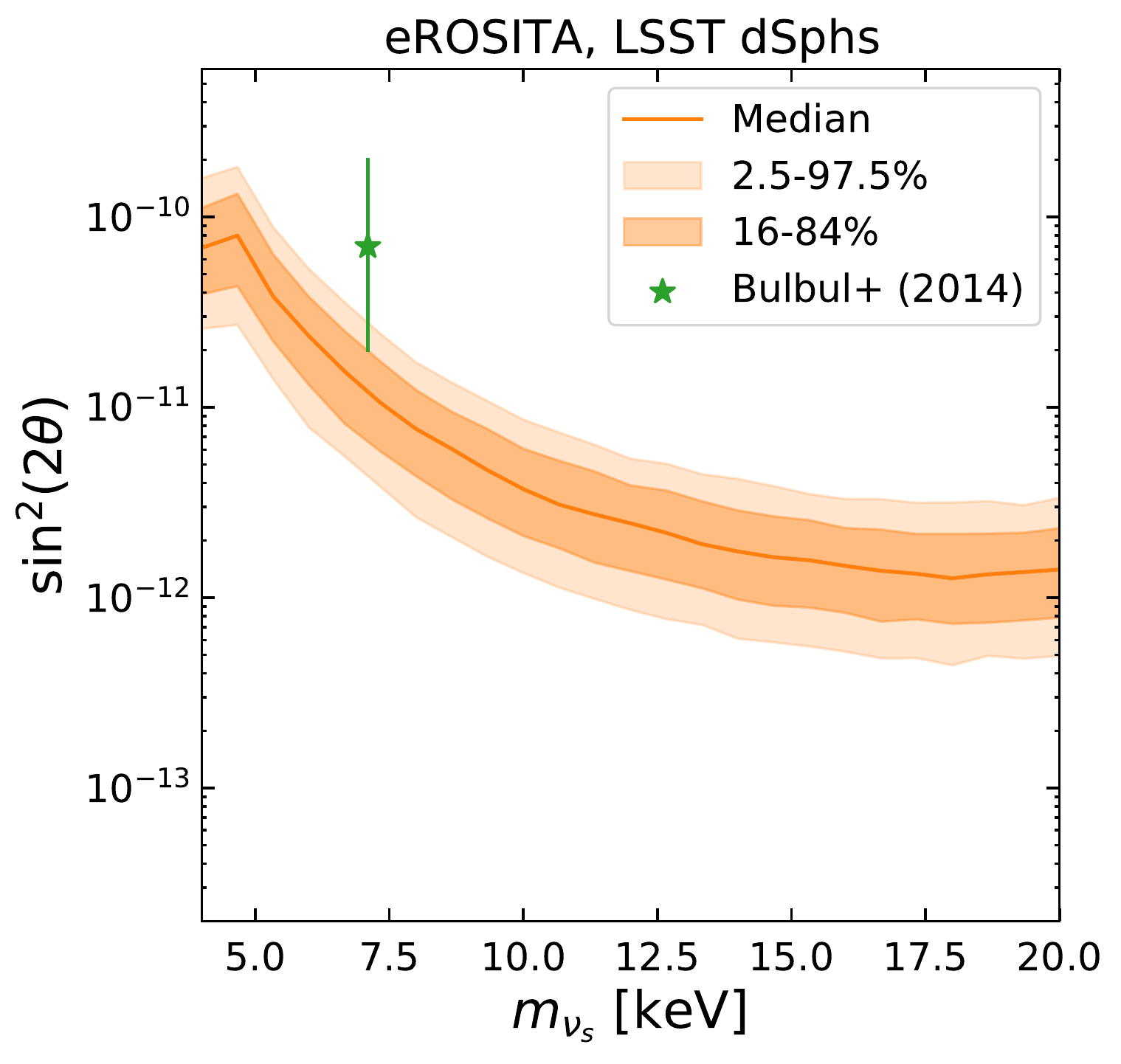}
    \caption{Left: Mixing angle constraints for eROSITA using $D$-factor data from all known dwarfs. Right: Mixing angle constraints for eROSITA using prospective all-sky LSST sources. In both panels, the analysis is performed for a single annulus of 0.5$^\circ$. For each mass, 500 Monte Carlo runs were performed.}
    \label{fig:groupB_eROSITA_annuli_compare}
\end{figure*}

For eROSITA that enables all-sky survey, we consider both known dSphs and potential LSST sources.
We perform a joint-likelihood analysis of all available dSphs using an integration angle of $\alpha_{\rm int} = 0.5^\circ$.
The subscript $i$ in Eq.~(\ref{eq:likelihood}) now runs over each dSph in the sample as well as energy bin.
In Fig.~\ref{fig:groupB_eROSITA_annuli_compare}, we show expected sensitivity to the mixing angle $\sin^22\theta$ for a given mass $m_{\nu_s}$, expected with all the known and future LSST dSphs.
These show that both the known and future LSST dSphs can be used to test the claim of 7~keV sterile neutrino~\cite{Bulbul:2014ala}, although the X-ray photon counts might fluctuate upward to prevent the solid conclusion.
Since the angular resolution of eROSITA is much better than the aperture of 0.5$^\circ$ that we considered here, we also repeated the same analysis but by dividing the regions of interest around each dSph into ten equal-width annuli.
We however find that this hardly changes the quantitative conclusion.
This is because the characteristic of the signal and the background is very different already in their energy distributions, and hence, adding one more degree of discriminating them does not help improve the eROSITA's sensitivity any further.

\section{Heavy dark matter}
\label{sec:Heavy dark matter}

For heavy dark matter, we consider candidates $\chi$ decaying into $b\bar{b}$ or $\tau^+\tau^-$ channels. Once a heavy dark matter particle decays, the daughter particles hadronize and one of the byproducts is gamma rays that can be detected with the gamma-ray observatories such as HAWC and CTA. We here consider the masses of these dark matter candidates to be in the range of 200~TeV--20~PeV. To calculate the expected gamma-ray event spectrum for a given source, we use the {\tt PPPC4DMID} numerical package~\cite{Ciafaloni:2010ti} for calculating the particle spectrum of dark matter decay for aforementioned channels with a maximum possible dark matter mass of 200~TeV. For masses above 200~TeV, we use the energy scaling proposed in Ref.~\cite{Chianese:2019kyl}:
\begin{equation}
     \frac{dN}{dE} = \frac{m_A}{m_\chi} \frac{dN_A}{dE'}\;\;\;(E' = E\frac{m_{A}}{m_{\chi}}),
\end{equation}
where $m_A = 200$~TeV is the reference mass and ${dN_A}/{dE}$ is the spectrum calculated at $m_A$. The obtained spectrum is used to calculate the gamma-ray flux from dark matter decay using Eq.~(\ref{eq:dFdE}).
We then use the gamma-ray flux to calculate the expected amount of events for an energy bin of a detector using Eq.~(\ref{eq:event counts (N)}). This section targets the detection of gamma-ray flux for heavy dark matter searches within the gamma-ray energy range between 300~GeV and 100~TeV.

A major background contribution to gamma-ray detection in current telescopes comes from cosmic rays, specifically from energetic electrons ($e$) and protons ($p$).
These fluxes are given as \cite{Silverwood:2014yza}
\begin{equation}\label{eq:e_flux}
    \frac{d^2\phi_\text{e}}{dEd\Omega} = 1.17\times 10^{-11}\left(\frac{E}{\text{TeV}}\right)^{-\Gamma}(\text{GeV}\;\text{cm}^{2}\;\text{s}\;\text{sr})^{-1},
\end{equation}
\begin{equation}\label{eq:h_flux}
    \frac{d^2\phi_\text{p}}{dEd\Omega} = 8.73\times 10^{-9}\left(\frac{E}{\text{TeV}}\right)^{-2.71}(\text{GeV}\;\text{cm}^{2}\;\text{s}\;\text{sr})^{-1},
\end{equation}
with
\begin{equation}
  \Gamma = \begin{cases}
         3.0, E < 1\;\text{TeV}, \\
         3.7, E > 1\;\text{TeV},
         \end{cases}   
\end{equation}
with which the aforementioned procedure for calculating the events per energy bin per given solid angle are performed to evaluate the background contributions. For different observatories, we consider the relevant hadron efficiencies which are then corrected against the hadron background flux during our analysis. The hadron rejection factor along with the effective area (with G/H cut) for HAWC has been extracted from Ref.~\cite{Goodman:HAWC}. For CTA, we have considered an overall cutoff factor of $10^{-2}$ while shifting the energies by a pre-factor of 3 in order to account for the reduced Cherenkov light emitted by hadronic showers~\cite{Fegan:1997db}. The effective areas for CTA telescopes are similarly extracted from \url{www.cta-observatory.org}. The relevant specifications further considered for corresponding telescopes are given in Table~\ref{tab:groupD_table_telescope_params}.

\begin{table}
    \scriptsize
	\caption{Summary of specifications considered for HAWC and CTA.}
	\begin{tabular}{ l c c c c c c }
		\hline
		\hline
		& HAWC & CTA\\
		\hline
		Flat energy resolution & 100\% & 10\% \\
		Minimal angular resolution & $0.5^\circ$ & $0.05^\circ$\\
		Exposure time\footnote{To each source.} & 2 years & 500 hours~\cite{CTAConsortium:2018tzg} \\
		Field of view (FOV)\footnote{CTA comprises multiple telescopes with different FOVs, and its overall FOV depends on the array layout. HAWC is an all-sky telescope.} & 2/3 of sky & 4.5--10$^\circ$ \\
		\hline
		\hline
	\end{tabular}
\label{tab:groupD_table_telescope_params}
\end{table}

\begin{figure*}[htbp!]
     \centering
     \includegraphics[height=5.5cm]{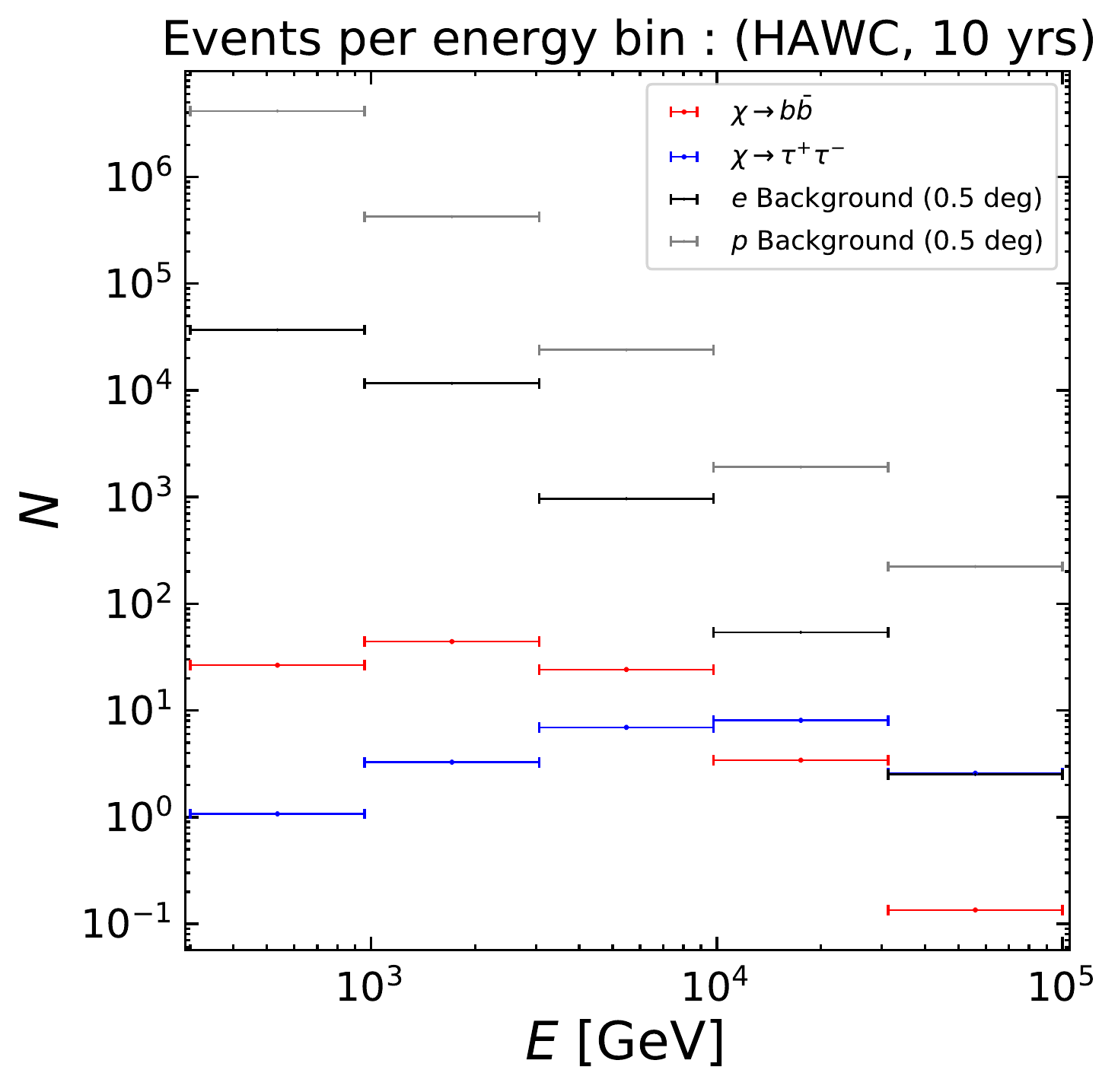}
     \includegraphics[height=5.5cm]{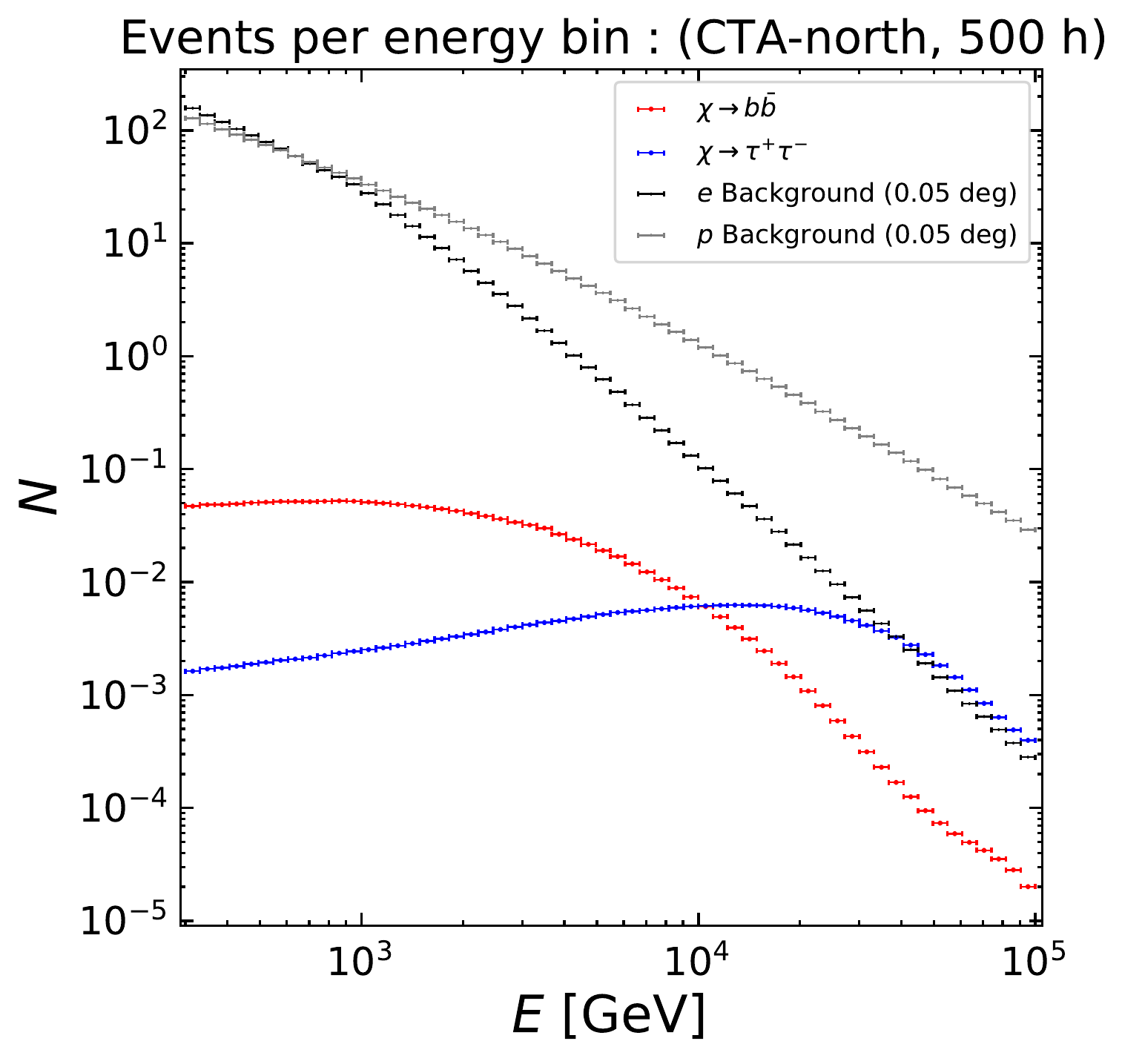}
     \includegraphics[height=5.5cm]{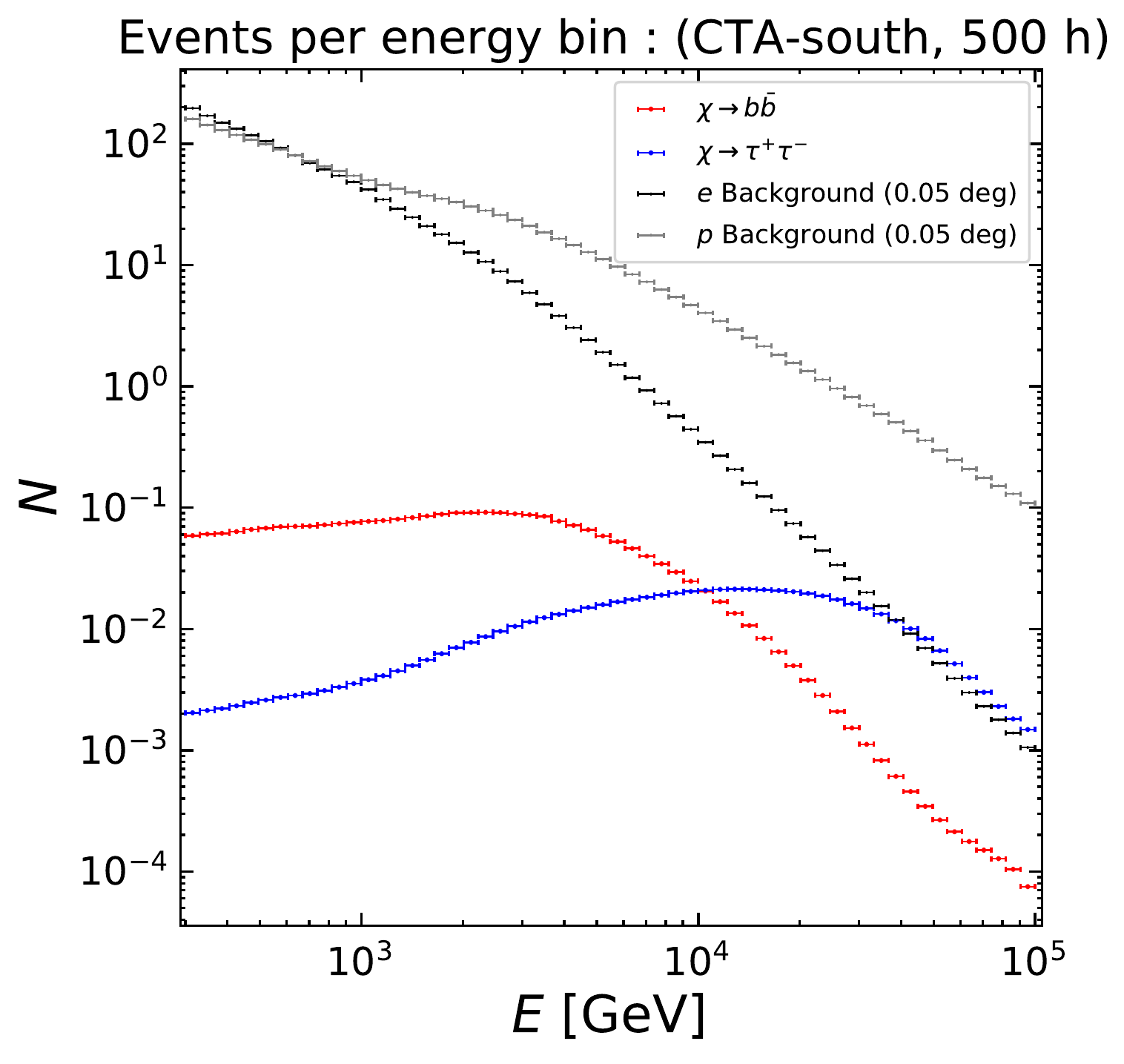}
     \caption{Event counts per energy bin, proton and electron backgrounds for HAWC ($\alpha_{\rm int} = 0.5^\circ$), CTA-north and CTA-south ($\alpha_{\rm int} = 0.05^\circ$), while considering efficiency cutoffs for the hadron background contributions for each telescope. The dark matter parameters adopted are $m_\chi$ = 200~TeV, $\Gamma_\chi = 10^{-28}~\text{s}^{-1}$, and $D = 10^{19}~\text{GeV}~\text{cm}^{-2}$ (HAWC) and $D = 10^{17}~\text{GeV}~\text{cm}^{-2}$ (CTA).}
     \label{fig:groupD_events}
\end{figure*}

In Fig.~\ref{fig:groupD_events}, we have presented gamma-ray event spectrum along with background events expected for each detector for a mock dark matter source with the $D$ factors of $10^{19}$ and $10^{17}$~GeV~cm$^{-2}$ for HAWC ($\alpha_{\rm int} = 0.5^\circ$) and CTA ($\alpha_{\rm int} = 0.05^\circ$), respectively, $m_\chi = 200$~TeV, and $\Gamma_\chi = 10^{-28}~\text{s}^{-1}$.

In order to project detector sensitivities on dark matter lifetime, we adopt a similar procedure as adopted in the previous section for identifying mixing angle limits for sterile neutrino dark matter, by evaluating the likelihood and TS. In this particular case, signal and background counts ($\mu_{i}(\Gamma)$) for a range of dark matter decay rates ($\Gamma_\chi$) at given dark matter mass ($m_\chi$) are calculated for each telescope as discussed above. We used the flat energy resolutions for these telescopes when calculating energy bins; i.e., the energy bin widths considered are constant in logarithm space. The next subsections present the obtained sensitivity limits of the considered heavy dark matter from both the HAWC and CTA.

\subsection{Results for HAWC}

In this subsection, we take 21 known dSphs, whose declination $\delta$ is within $\pm 45^\circ$ around the position of the HAWC, $19^\circ$~\cite{Abeysekara:2011yu} such that they are within HAWC's field of view.
We correct for the exposure of the HAWC that is the largest toward the zenith, by multiplying it by an extra factor of $\cos(\delta-19^\circ)$.
As it mainly surveys the northern sky, we do not discuss implications from the future LSST dSphs that will be mostly found in the southern sky.
We consider a single detection annulus of $0.5^\circ$ degree, corresponding to HAWC's minimal angular resolution.
An extra step while finding the joint likelihood is to consider a sum of all log-likelihoods obtained from each candidates (dSphs) :
\begin{equation}\label{eq:comb_like}
    \ln\mathcal{L}(\Gamma) = \sum_{\rm dSph}
    \ln\mathcal{L}_{\rm dSph}(\Gamma) ,
\end{equation}
from which the lower limits of decay lifetime with 95\% CL, $\tau_{95}$, are obtained similarly as in Sec.~\ref{sec:sterile_neutr_DM}. For each dark matter mass considered, we have obtained 1000 $\tau_{95}$ values to remedy any statistical anomalies in our computations.

\begin{figure*}[htbp!]
     \centering
     \includegraphics[width=8cm]{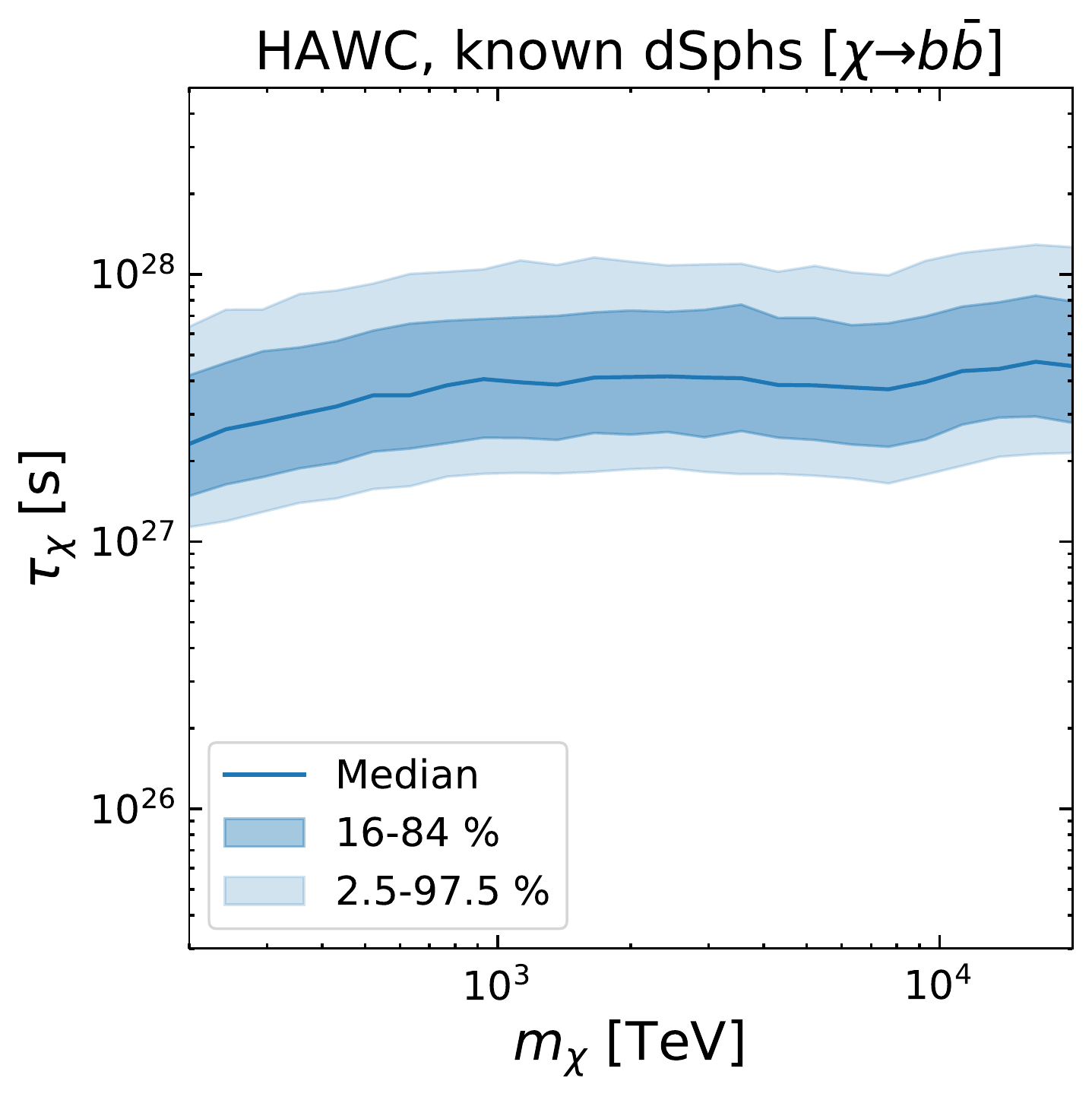}
     \includegraphics[width=8cm]{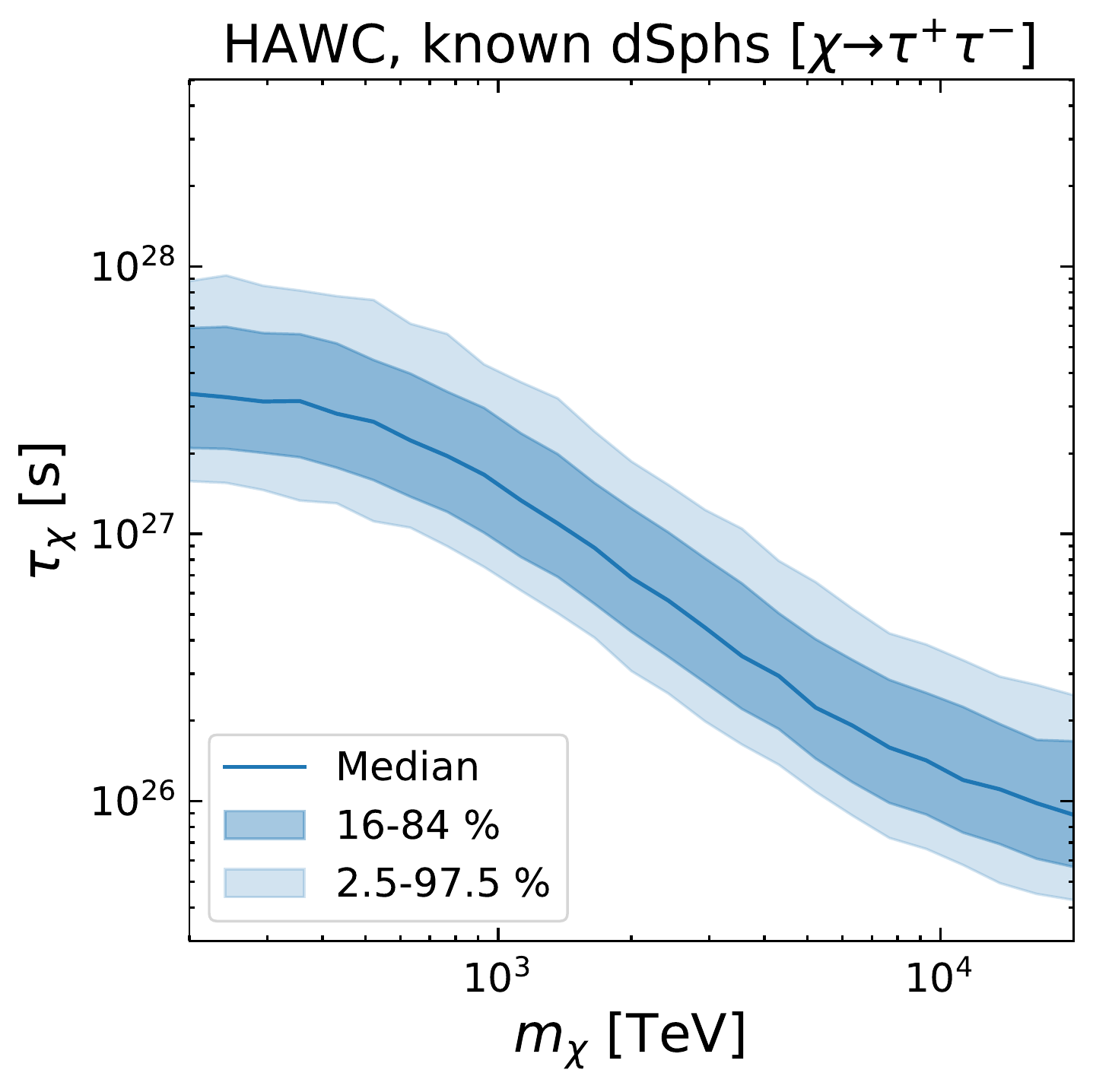}
     \caption{Lower limits on dark matter decay lifetime from the joint likelihood analysis of 21 known dSphs that are within the field of view of HAWC. The constraints are given for both $\chi \to b\bar{b}$ channel (left) and $\chi \to \tau^+\tau^-$ channel (right). The median values are depicted as a solid line, and the 68\% and 95\% containment bands are shown in thick and thin relative shadings, respectively, as a result of the 1000 runs of obtaining mock data.}
     \label{fig:groupD_gm_hawc}
\end{figure*}

In Fig.~\ref{fig:groupD_gm_hawc}, we present the results for both $b\bar b$ and $\tau^+\tau^-$ decay channels for the considered known 21 dSphs. Individual likelihood analysis for each of the known dSphs is presented in Appendix (Fig.~\ref{fig:groupA_HAWC}) presenting only the median values of the distribution of 1000 $\tau_{95}$ values obtained per dark matter mass. 
The HAWC's sensitivity is stronger than $10^{27}$~s nearly independent of mass for the $b\bar b$, while it gets reduced to $10^{26}$~s for the $\tau^+\tau^-$ if dark matter mass is heavier than $\sim$10~PeV.
This is because for dark matter in this mass range, most of the gamma-ray photons are emitted well above the energy range that we consider, as the spectrum for the $\tau^+\tau^-$ channel is much harder than the case of the $b\bar b$ channel.

\subsection{Results for CTA}

\begin{figure*}[htbp!]
    \centering
    \includegraphics[width=8cm]{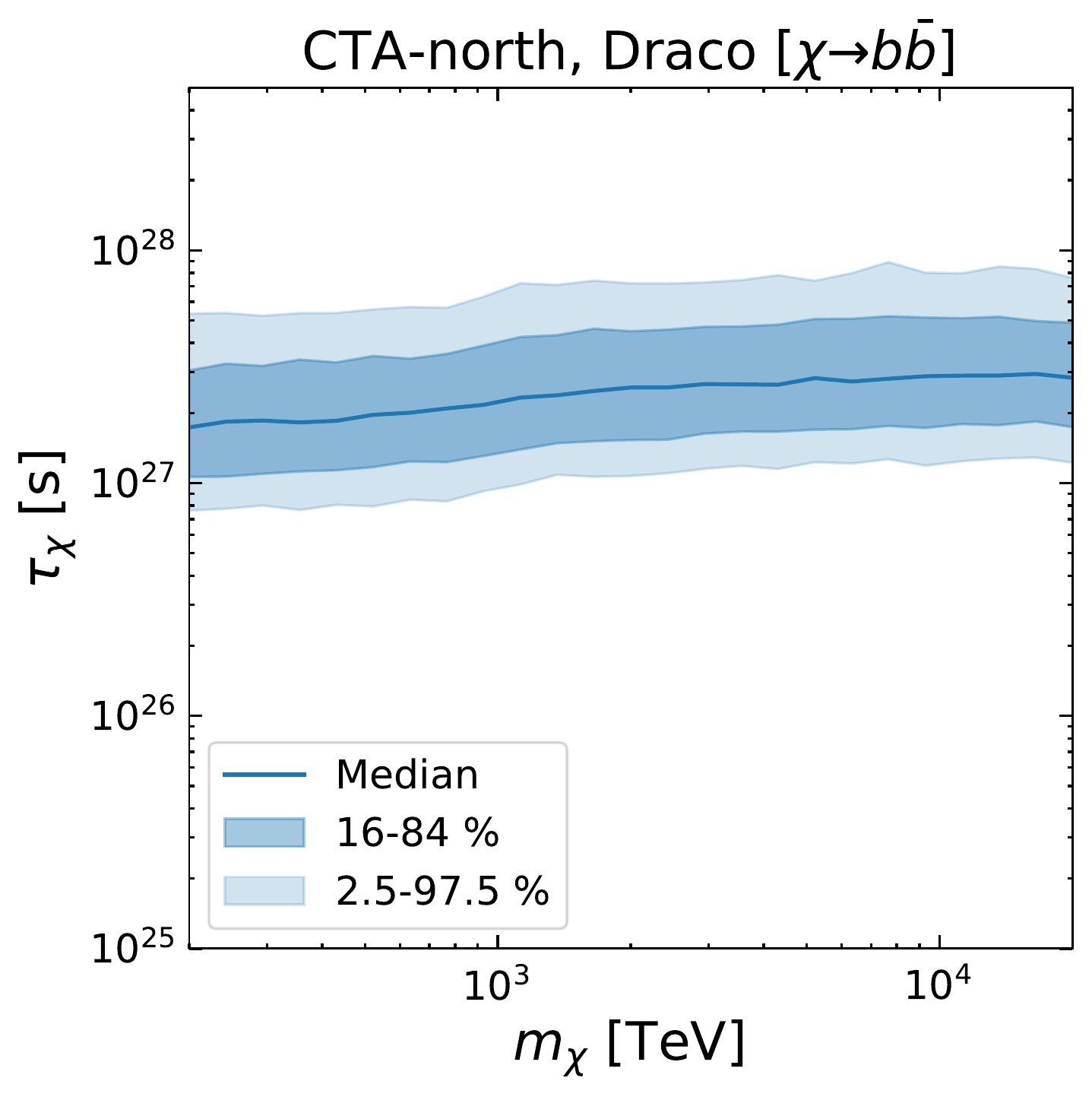}
    \includegraphics[width=8cm]{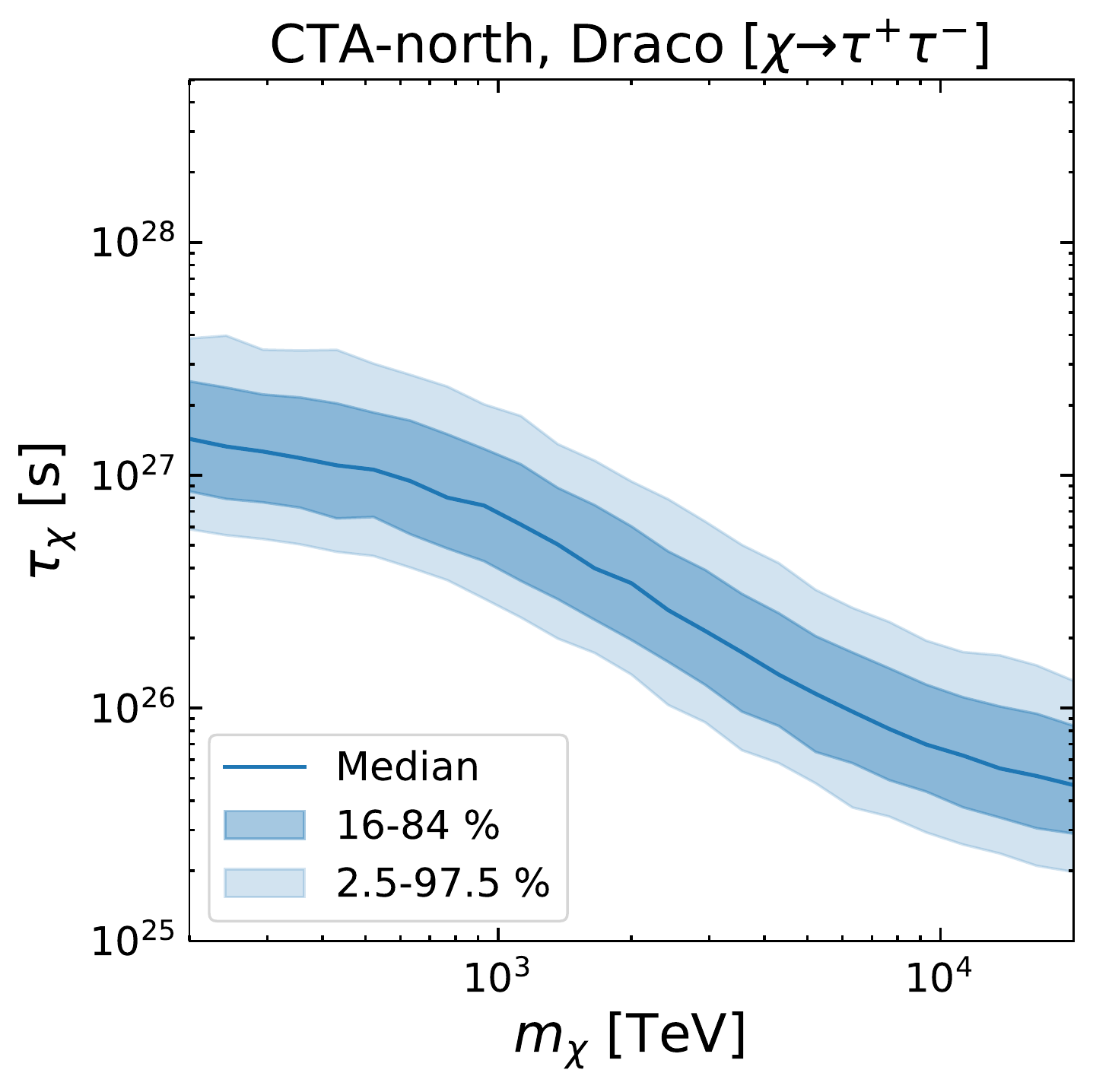}
    \includegraphics[width=8cm]{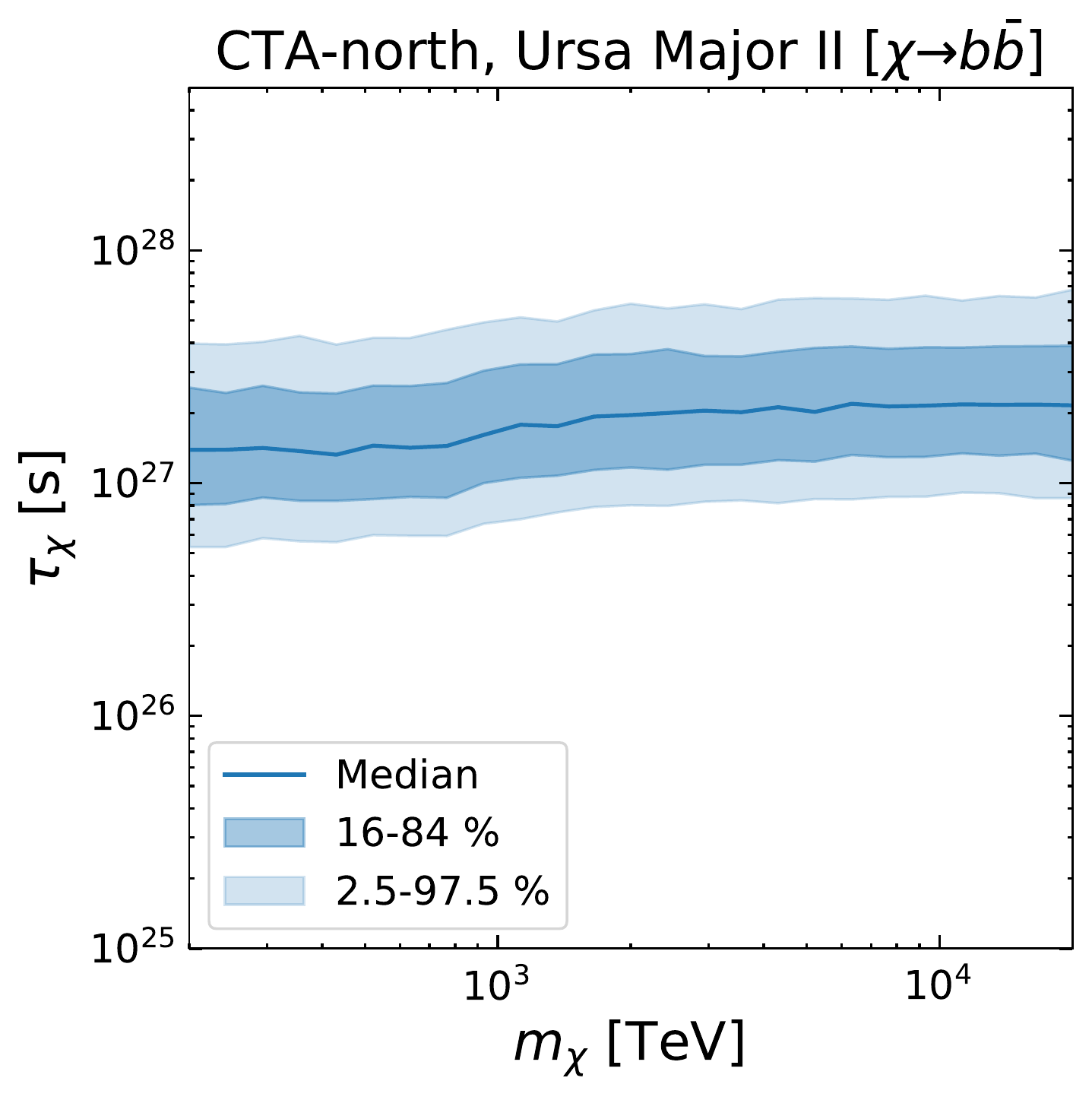}
    \includegraphics[width=8cm]{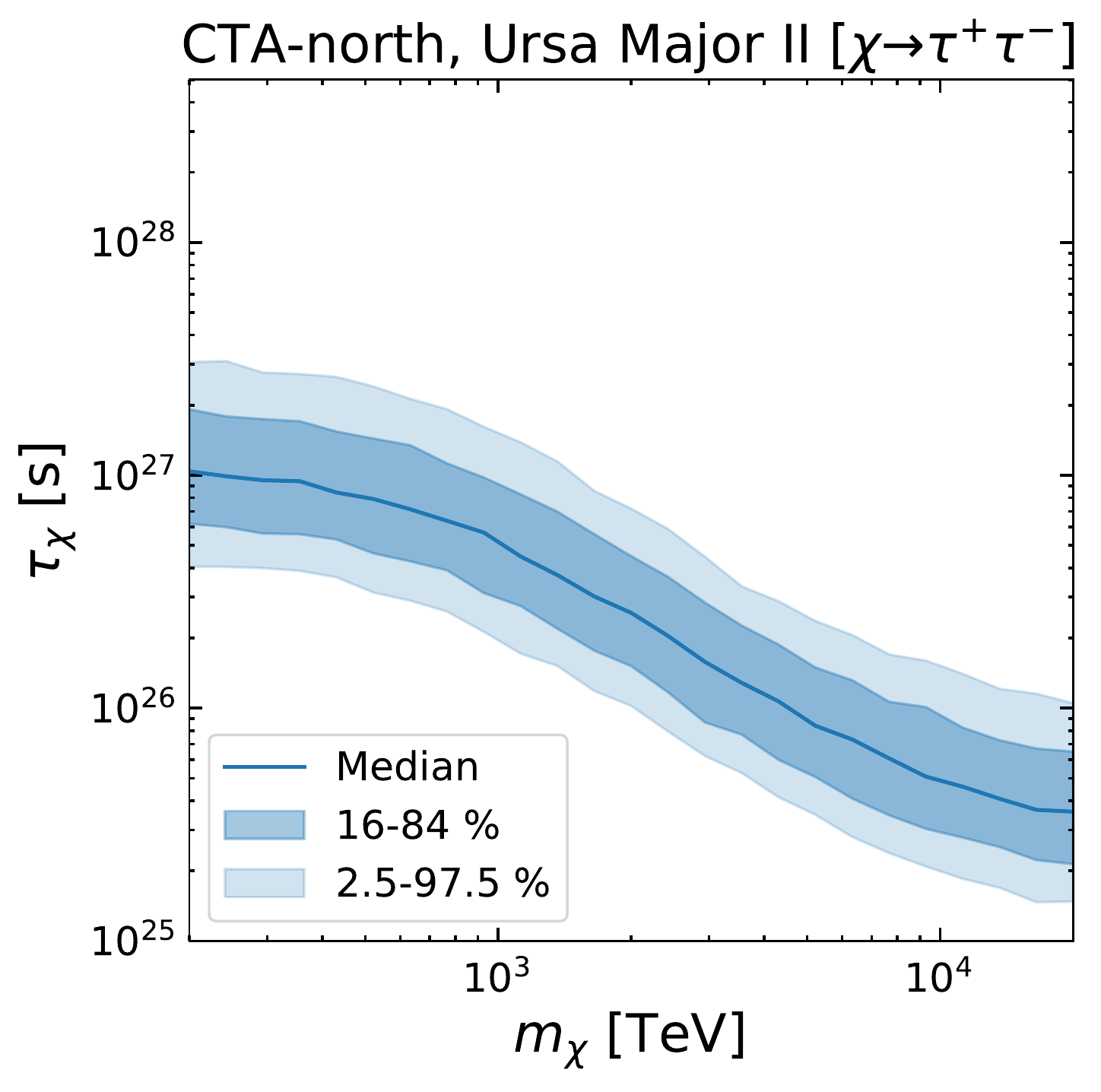}
    \caption{Dark matter lifetime constraints for some the best known dSphs, Draco (top) and Ursa Major II (bottom), detectable with the CTA North telescopes, for the $b\bar b$ (left) and $\tau^+\tau^-$ (right) final states. The solid lines show the medians, while the light and dark bands show 68\% and 95\% containment intervals, respectively, as a result of the 1000 Monte Carlo simulations of mock data.}
    \label{fig:groupD_gm_cta_north_single}
\end{figure*}

CTA will have much better angular resolution than HAWC. 
Taking advantage of this, within its field of view of $\sim$5--10$^\circ$, one can have multiple annuli.
Here we adopt 10 annuli, each of whose widths is 0.05$^\circ$ within the region with radius of 0.5$^\circ$.
However, since the field of view is limited to a narrow region of sky, we mainly focus on the few best target dSphs.
Among the known dSphs, we discuss one classical dSph, Draco, and one ultrafaint dSph, Ursa Major II.
Figure~\ref{fig:groupD_gm_cta_north_single} shows the expected constraints on the dark matter lifetime with Draco and Ursa Major II, simulated for the CTA North for all both the $b\bar b$ and $\tau^+ \tau^-$ decay channels. Here, we have again considered $1000$ $\tau_{95}$ values per dark matter mass while computing the sensitivity bands.
We find that the generic trend of these limits is the same as those obtained for HAWC, but the limits are overall slightly weaker.
This is because of combined statistical power of all dSphs that can be seen with HAWC thanks to its large field of view, albeit its less good energy and angular resolution.

\begin{figure*}[htbp!]
    \centering
    \includegraphics[width=8cm]{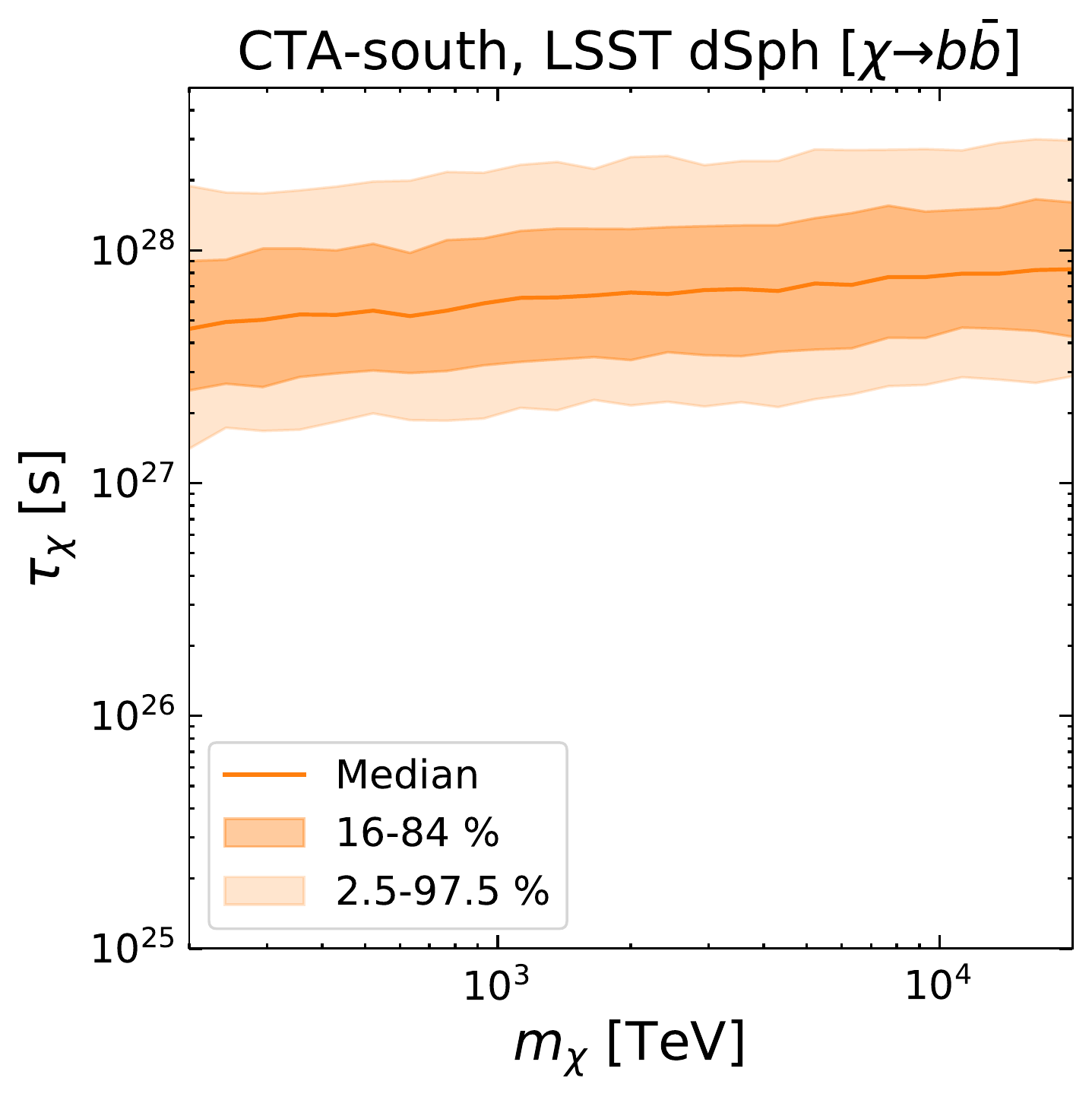}
    \includegraphics[width=8cm]{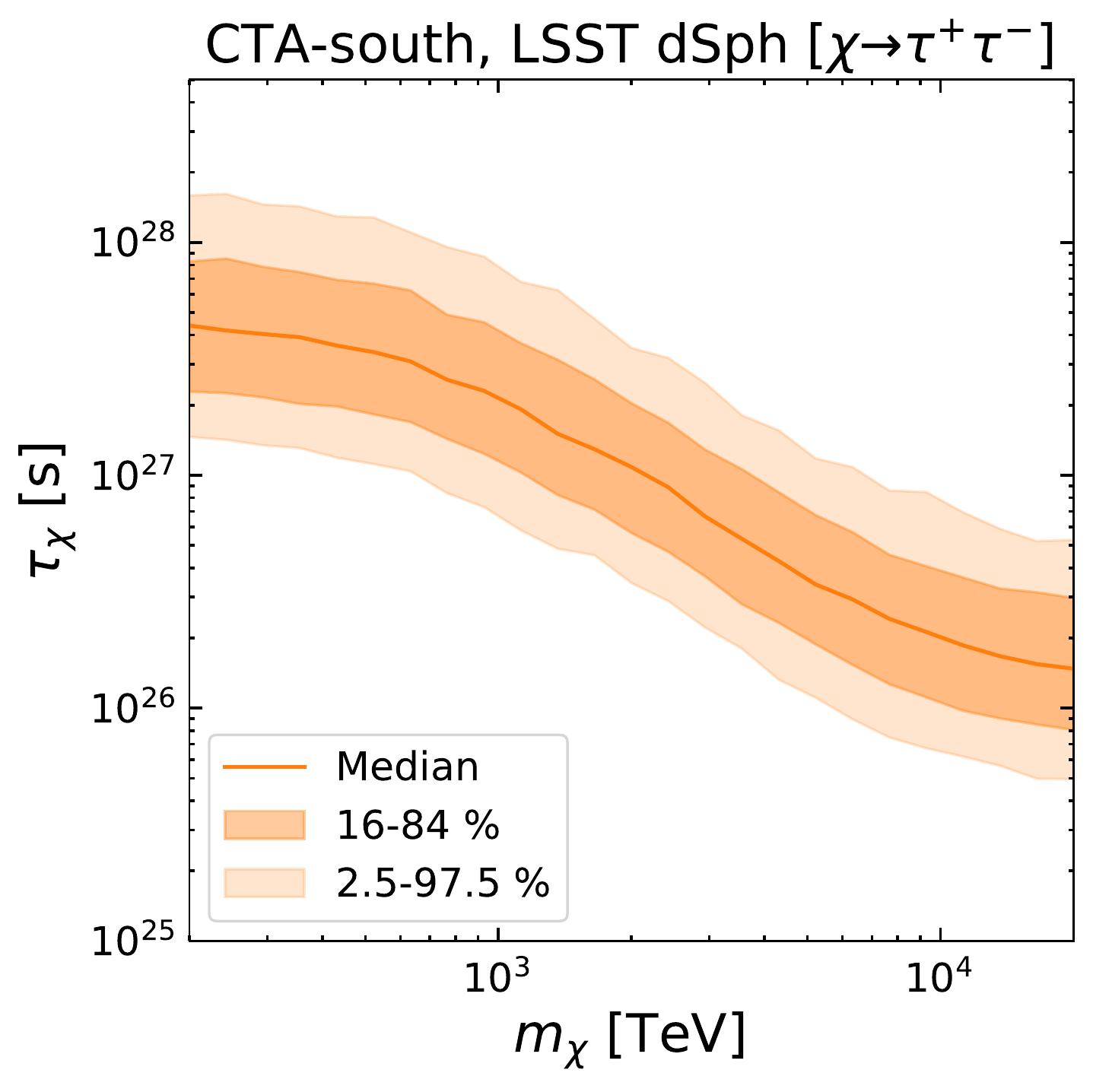}
    \caption{The same as Fig.~\ref{fig:groupD_gm_cta_north_single}, but for a dSph with the largest $D$ factor that would be detected with the LSST.}
    \label{fig:groupD_gm_cta_south_single}
\end{figure*}

CTA South will survey the entire southern sky, which will allow us to observe dSphs detected with LSST. We consider through Monte Carlo simulations the best possible LSST dSph candidate that would possess the largest $D$ factor from their host subhalo.
The results are shown in Figure~\ref{fig:groupD_gm_cta_south_single}. 
The expected limits are stronger by a factor of a few than the case of Draco and Ursa Major II that can be observed with the CTA North.

\section{Discussion and conclusions}
\label{sec:Conclusions}

In this paper, we made a comprehensive study of dark matter decays in the dwarf spheroidal galaxies.
By revisiting their density profile estimates with the latest models of subhalo and satellite formation, we updated predictions of X-ray and gamma-ray fluxes from dwarf galaxies.
We then discussed detectability of dark matter signals at X-ray telescopes (eROSITA, XRISM, and Athena), if dark matter is made of sterile neutrinos with keV masses, and high-energy gamma-ray observatories (HAWC and CTA) if dark matter is made of metastable particles heavier than conventional WIMPs.

With physically motivated priors, we show that the posterior distributions of the astrophysical $D$-factors for dSphs are narrower, but are also typically lower compared to the case with uninformative prior. 
Such an effect was shown for the case of dark matter annihilation, but to a stronger degree~\cite{Ando:2020yyk}. With our revised estimates of the $D$-factors, our results are more robust, though also weaker than previously expected.

For pointing X-ray satellites, XRISM and Athena, we performed sensitivity estimates for two of the most promising dwarfs: Draco (classical) and Ursa Major II (ultrafaint).
With both these instruments, we should be able to critically test much debated 3.5-keV X-ray lines, which suggested mixing angle of $\sin^22\theta = 7\times 10^{-11}$ for 7.1~keV sterile neutrinos.  The excellent energy resolution of these detectors will also allow the use of velocity spectroscopy as a detection diagnostic tool~\cite{Speckhard:2015eva, Powell:2016zbo, Zhong:2020wre}. 
For the all-sky instrument, eROSITA, we are able to combine information of all the dwarfs. 
It is also sensitive to test the claimed 3.5-keV line for the current known dwarfs and also tens of ultrafaint dwarf galaxies that might be found with the LSST in the future.

The dwarf galaxies are among many other complementary target regions.
Besides galaxies and clusters of galaxies that are already studied extensively~\cite{Bulbul:2014ala, Boyarsky:2014jta}, the cosmic X-ray background~\cite{Dessert:2018qih, Foster:2021ngm} and their anisotropies and cross correlations~\cite{Zandanel:2015xca, Caputo:2019djj} have also been proven to be efficient in constraining sterile neutrino dark matter.
It would also be interesting to study the expected sensitivity of eROSITA with the Galactic halo analysis, by adopting, e.g., inner regions around the Galactic center (A.~Dekker et al., in preparation).

For heavy dark matter, we explore the mass range between 200~TeV and 20~PeV that decays into $b\bar b$ or $\tau^+\tau^-$.
We are able to probe the lifetime of the heavy dark matter in the range of $10^{27}$--$10^{28}$~s for the $b\bar{b}$ channel, whereas the limits are weaker for $\tau^+\tau^-$ channel for mass ranges greater than PeV. Both CTA and HAWC will yield similar sensitivity to the dark matter decay. Although HAWC has smaller effective area than CTA, it has a larger field of view that enables greater sky coverage. This enables joint likelihood analysis of all the dwarfs in its field of view, compensating its smaller effective area. In the future, once LHAASO is completed, it is expected to be more sensitive than HAWC due to its larger effective area and broader energy coverage.

If heavy dark matter decays to either $b\bar{b}$ or $\tau^+ \tau^-$ as was investigated in this study, these final states create a number of lower energy photons through hadronic and electromagnetic cascades.
This was studied in a broader multimessenger context in Refs.~\cite{Murase:2015gea, Cohen:2016uyg, Ishiwata:2019aet}, which show comparable lower limits to the dark matter lifetime on the order of $10^{28}$~s or so.
In addition, it is expected that future neutrino telescopes such as IceCube-Gen2 and KM3NeT are able to probe similar parameter regions~\cite{Dekker:2019gpe, Ng:2020ghe}.
Therefore, both the HAWC and CTA will provide complementary constraints using other messengers.

\acknowledgments

SA acknowledges the support by JSPS/MEXT KAKENHI Grant Numbers JP17H04836, JP18H04340, JP20H05850, and JP20H05861 (SA).
KCYN was supported by the European Union’s Horizon 2020 research and innovation programme under the Marie Skłodowska-Curie grant agreement No 844664 (KCYN)
This work was carried out over the two-month period in April and May 2020 in the MSc course, ``Project Academic Skills for Research'' at the University of Amsterdam.
All the tasks were distributed to four sub-groups that worked on (a) $D$-factor estimates of known dwarfs (AGC, YM, INM); (b) $D$-factor estimates of LSST dwarfs (JK, MSPM, EP); (c) sterile neutrino constraints (ZF, SK, NMDS, FZ); and (d) heavy dark matter constraints (SKB, MF, CAvV), respectively, supervised by SA and KCYN.

\appendix

\section{Dark matter decay in known dwarf galaxies}
\label{app:Summary of dark matter decay in known dwarf galaxies}

\begin{table}
\caption{$D$-factor medians and corresponding 68\% credible intervals of the classical dSphs for $\alpha_{\rm int}=0.5^{\circ}$. The third column ($R_{D_{\rm MW}}$) represents the relative contribution of the Milky Way to the total $D$-factor: $R_{D_{\rm MW}} = D_{\rm MW}/(D_{\rm MW}+D_{\rm dSph})$.}
\begin{tabular}{ c c c }
\hline
\hline
Name & Classical & $R_{D_{\rm MW}}$ \\ 
\hline
Sagittarius & $18.59_{-0.09}^{+0.10}$ & 0.81 \\ 
Draco & $18.30_{-0.21}^{+0.18}$ & 0.68 \\ 
Ursa Minor & $18.26_{-0.18}^{+0.15}$ & 0.67 \\ 
Sculptor & $18.22_{-0.20}^{+0.16}$ & 0.71 \\ 
Fornax & $18.06_{-0.11}^{+0.09}$ & 0.76 \\ 
Sextans & $17.96_{-0.23}^{+0.21}$ & 0.79 \\ 
Leo I & $17.76_{-0.28}^{+0.23}$ & 0.85 \\ 
Carina & $17.69_{-0.37}^{+0.29}$ & 0.88 \\ 
Leo II & $17.58_{-0.38}^{+0.32}$ & 0.90 \\ 
\hline
\hline
\end{tabular}
\label{tab:groupA_table_classicals_0.5}
\end{table}

\begin{table}
\caption{The same as Table~\ref{tab:groupA_table_classicals_0.5}, but for $\alpha_{\rm int}=0.05^{\circ}$.}
\begin{tabular}{ c c c }
\hline
\hline
Name & Classical & $R_{D_{\rm MW}}$ \\ 
\hline
Sagittarius & $ 16.98_{-0.15}^{+0.15}$ & 0.64 \\ 
Draco & $ 16.87_{-0.11}^{+0.10}$ & 0.36 \\ 
Ursa Minor & $ 16.84_{-0.10}^{+0.10}$ & 0.35 \\ 
Sculptor & $ 16.82_{-0.10}^{+0.09}$ & 0.39 \\ 
Fornax & $ 16.74_{-0.09}^{+0.10}$ & 0.39 \\ 
Sextans & $ 16.68_{-0.12}^{+0.11}$ & 0.41 \\ 
Leo I & $ 16.62_{-0.05}^{+0.04}$ & 0.43 \\ 
Carina & $ 16.56_{-0.14}^{+0.12}$ & 0.51 \\ 
Leo II & $ 16.51_{-0.07}^{+0.07}$ & 0.52 \\ 
\hline
\hline
\end{tabular}
\label{tab:groupA_table_classicals_0.05}
\end{table}

\begin{table*}
\caption{$D$-factor medians and 68\% credible intervals of the ultrafaint dSphs for $\alpha_{\rm int}=0.5^{\circ}$ and the different values of $V_{\rm peak}$. $R_{D_{\rm MW}}$ represents the relative contribution of the Milky Way to the total $D$-factor: $R_{D_{\rm MW}} = D_{\rm MW}/(D_{\rm MW}+D_{\rm dSph})$.}
\begin{tabular}{ c c c c c c c }
\hline
\hline
Name & $V_{\rm peak}>14~\rm km~s^{-1}$ & $R_{D_{\rm MW}}$ & $V_{\rm peak}>18~\rm km~s^{-1}$ & $R_{D_{\rm MW}}$ & $V_{\rm peak}>22~\rm km~s^{-1}$ & $R_{D_{\rm MW}}$ \\ 
\hline
Segue 1 & $18.27_{-0.34}^{+0.22}$ & 0.63 & $18.32_{-0.33}^{+0.21}$ & 0.60 & $18.40_{-0.35}^{+0.21}$ & 0.56 \\ 
Ursa Major II & $18.25_{-0.26}^{+0.20}$ & 0.62 & $18.29_{-0.25}^{+0.18}$ & 0.60 & $18.38_{-0.26}^{+0.16}$ & 0.54 \\ 
Draco II & $18.08_{-0.47}^{+0.32}$ & 0.76 & $18.16_{-0.47}^{+0.29}$ & 0.73 & $18.29_{-0.42}^{+0.28}$ & 0.66 \\ 
Reticulum II & $18.07_{-0.39}^{+0.25}$ & 0.77 & $18.11_{-0.39}^{+0.26}$ & 0.75 & $18.17_{-0.41}^{+0.28}$ & 0.73 \\ 
Coma Berenices & $18.03_{-0.30}^{+0.21}$ & 0.79 & $18.06_{-0.31}^{+0.22}$ & 0.77 & $18.15_{-0.52}^{+0.30}$ & 0.74 \\ 
Ursa Major I & $17.96_{-0.21}^{+0.20}$ & 0.77 & $18.00_{-0.54}^{+0.32}$ & 0.75 & $18.11_{-0.33}^{+0.25}$ & 0.70 \\ 
Tucana II & $17.86_{-0.54}^{+0.31}$ & 0.89 & $17.94_{-0.52}^{+0.29}$ & 0.88 & $18.07_{-0.55}^{+0.34}$ & 0.84 \\ 
Triangulum II & $17.84_{-0.63}^{+0.40}$ & 0.81 & $17.91_{-0.60}^{+0.39}$ & 0.78 & $18.07_{-0.47}^{+0.28}$ & 0.71 \\ 
Bootes II & $17.82_{-0.69}^{+0.38}$ & 0.89 & $17.90_{-0.63}^{+0.36}$ & 0.87 & $18.06_{-0.25}^{+0.16}$ & 0.82 \\ 
Carina II & $17.74_{-0.43}^{+0.34}$ & 0.88 & $17.78_{-0.43}^{+0.36}$ & 0.87 & $17.86_{-0.69}^{+0.37}$ & 0.85 \\ 
Hyrdus 1 & $17.65_{-0.32}^{+0.30}$ & 0.92 & $17.71_{-0.73}^{+0.37}$ & 0.91 & $17.79_{-0.46}^{+0.37}$ & 0.90 \\ 
Horologium I & $17.64_{-0.74}^{+0.38}$ & 0.90 & $17.64_{-0.34}^{+0.30}$ & 0.90 & $17.62_{-0.39}^{+0.28}$ & 0.91 \\ 
Bootes I & $17.62_{-0.24}^{+0.24}$ & 0.93 & $17.62_{-0.25}^{+0.24}$ & 0.93 & $17.62_{-0.77}^{+0.42}$ & 0.93 \\ 
Canes Venatici I & $17.51_{-0.14}^{+0.15}$ & 0.93 & $17.52_{-0.15}^{+0.15}$ & 0.93 & $17.59_{-0.31}^{+0.23}$ & 0.92 \\ 
Aquarius 2 & $17.44_{-0.77}^{+0.40}$ & 0.95 & $17.50_{-0.76}^{+0.40}$ & 0.94 & $17.54_{-0.59}^{+0.51}$ & 0.94 \\ 
Canes Venatici II & $17.38_{-0.66}^{+0.36}$ & 0.94 & $17.41_{-0.67}^{+0.38}$ & 0.94 & $17.50_{-0.16}^{+0.20}$ & 0.93 \\ 
Segue 2 & $17.31_{-0.72}^{+0.54}$ & 0.93 & $17.36_{-0.67}^{+0.55}$ & 0.93 & $17.50_{-0.75}^{+0.44}$ & 0.90 \\ 
Grus I & $17.12_{-1.09}^{+0.62}$ & 0.98 & $17.25_{-1.06}^{+0.59}$ & 0.97 & $17.48_{-1.04}^{+0.54}$ & 0.95 \\ 
Eridanus II & $17.10_{-0.42}^{+0.31}$ & 0.97 & $17.17_{-1.05}^{+0.53}$ & 0.96 & $17.36_{-1.05}^{+0.53}$ & 0.94 \\ 
Pisces II & $17.07_{-1.06}^{+0.54}$ & 0.97 & $17.16_{-0.48}^{+0.29}$ & 0.97 & $17.31_{-0.27}^{+0.41}$ & 0.96 \\ 
Leo T & $17.02_{-0.60}^{+0.37}$ & 0.97 & $17.09_{-0.62}^{+0.34}$ & 0.96 & $17.28_{-1.04}^{+0.53}$ & 0.94 \\ 
Leo V & $16.99_{-1.05}^{+0.60}$ & 0.98 & $17.09_{-1.03}^{+0.58}$ & 0.97 & $17.27_{-1.04}^{+0.58}$ & 0.95 \\ 
Tucana III & $16.94_{-0.39}^{+0.54}$ & 0.98 & $17.06_{-1.08}^{+0.56}$ & 0.98 & $17.23_{-0.56}^{+0.36}$ & 0.97 \\ 
Hercules & $16.93_{-0.61}^{+0.54}$ & 0.99 & $17.05_{-0.35}^{+0.49}$ & 0.99 & $17.20_{-0.70}^{+0.36}$ & 0.98 \\ 
Pegasus III & $16.93_{-1.05}^{+0.59}$ & 0.98 & $16.87_{-0.59}^{+0.56}$ & 0.98 & $16.83_{-0.79}^{+0.78}$ & 0.99 \\ 
Leo IV & $16.78_{-0.87}^{+0.73}$ & 0.99 & $16.81_{-0.86}^{+0.74}$ & 0.98 & $16.75_{-0.53}^{+0.61}$ & 0.99 \\ 
\hline
\hline
\end{tabular}
\label{tab:groupA_table_UFD_0.5}
\end{table*}

\begin{table*}
\caption{The same as Table~\ref{tab:groupA_table_UFD_0.5} but for $\alpha_{\rm int}=0.05^{\circ}$.}
\begin{tabular}{ c c c c c c c }
\hline
\hline
Name & $V_{\rm peak}=14~\rm km~s^{-1}$ & $R_{D_{\rm MW}}$ & $V_{\rm peak}=18~\rm km~s^{-1}$ & $R_{D_{\rm MW}}$ & $V_{\rm peak}=22~\rm km~s^{-1}$ & $R_{D_{\rm MW}}$ \\ 
\hline
Segue 1 & $16.77_{-0.18}^{+0.16}$ & 0.35 & $16.79_{-0.17}^{+0.15}$ & 0.34 & $16.85_{-0.17}^{+0.14}$ & 0.31 \\ 
Ursa Major II & $16.75_{-0.17}^{+0.15}$ & 0.34 & $16.77_{-0.15}^{+0.14}$ & 0.33 & $16.83_{-0.15}^{+0.13}$ & 0.30 \\ 
Ursa Major I & $16.65_{-0.12}^{+0.11}$ & 0.41 & $16.66_{-0.23}^{+0.21}$ & 0.40 & $16.77_{-0.22}^{+0.18}$ & 0.34 \\ 
Coma Berenices & $16.63_{-0.15}^{+0.13}$ & 0.48 & $16.66_{-0.11}^{+0.10}$ & 0.46 & $16.72_{-0.17}^{+0.15}$ & 0.43 \\ 
Draco II & $16.62_{-0.25}^{+0.21}$ & 0.48 & $16.65_{-0.14}^{+0.12}$ & 0.46 & $16.71_{-0.23}^{+0.19}$ & 0.43 \\ 
Reticulum II & $16.62_{-0.18}^{+0.16}$ & 0.49 & $16.65_{-0.18}^{+0.15}$ & 0.47 & $16.70_{-0.10}^{+0.09}$ & 0.44 \\ 
Tucana II & $16.54_{-0.26}^{+0.20}$ & 0.64 & $16.58_{-0.25}^{+0.19}$ & 0.62 & $16.68_{-0.22}^{+0.18}$ & 0.56 \\ 
Bootes II & $16.51_{-0.29}^{+0.23}$ & 0.62 & $16.56_{-0.29}^{+0.22}$ & 0.59 & $16.67_{-0.26}^{+0.20}$ & 0.53 \\ 
Triangulum II & $16.48_{-0.27}^{+0.22}$ & 0.49 & $16.53_{-0.26}^{+0.22}$ & 0.46 & $16.65_{-0.22}^{+0.20}$ & 0.39 \\ 
Carina II & $16.46_{-0.20}^{+0.16}$ & 0.59 & $16.49_{-0.29}^{+0.22}$ & 0.57 & $16.60_{-0.28}^{+0.19}$ & 0.51 \\ 
Horologium I & $16.44_{-0.31}^{+0.22}$ & 0.60 & $16.48_{-0.19}^{+0.16}$ & 0.58 & $16.53_{-0.17}^{+0.15}$ & 0.55 \\ 
Canes Venatici I & $16.41_{-0.05}^{+0.05}$ & 0.62 & $16.42_{-0.11}^{+0.10}$ & 0.61 & $16.47_{-0.28}^{+0.20}$ & 0.59 \\ 
Bootes I & $16.41_{-0.12}^{+0.10}$ & 0.67 & $16.41_{-0.12}^{+0.12}$ & 0.67 & $16.45_{-0.11}^{+0.10}$ & 0.65 \\ 
Hyrdus 1 & $16.39_{-0.12}^{+0.12}$ & 0.68 & $16.41_{-0.05}^{+0.05}$ & 0.67 & $16.44_{-0.10}^{+0.10}$ & 0.66 \\ 
Aquarius 2 & $16.34_{-0.30}^{+0.21}$ & 0.70 & $16.39_{-0.30}^{+0.20}$ & 0.68 & $16.42_{-0.27}^{+0.19}$ & 0.66 \\ 
Canes Venatici II & $16.33_{-0.25}^{+0.19}$ & 0.65 & $16.36_{-0.26}^{+0.18}$ & 0.63 & $16.42_{-0.05}^{+0.04}$ & 0.60 \\ 
Segue 2 & $16.23_{-0.23}^{+0.24}$ & 0.63 & $16.29_{-0.21}^{+0.22}$ & 0.60 & $16.40_{-0.16}^{+0.19}$ & 0.54 \\ 
Eridanus II & $16.22_{-0.17}^{+0.16}$ & 0.68 & $16.26_{-0.41}^{+0.29}$ & 0.66 & $16.39_{-0.36}^{+0.27}$ & 0.59 \\ 
Grus I & $16.19_{-0.43}^{+0.29}$ & 0.79 & $16.25_{-0.42}^{+0.26}$ & 0.77 & $16.36_{-0.39}^{+0.25}$ & 0.72 \\ 
Pisces II & $16.17_{-0.44}^{+0.28}$ & 0.75 & $16.24_{-0.17}^{+0.14}$ & 0.72 & $16.35_{-0.09}^{+0.11}$ & 0.66 \\ 
Leo T & $16.17_{-0.25}^{+0.20}$ & 0.67 & $16.21_{-0.26}^{+0.18}$ & 0.65 & $16.33_{-0.41}^{+0.25}$ & 0.58 \\ 
Leo V & $16.14_{-0.42}^{+0.29}$ & 0.74 & $16.20_{-0.44}^{+0.27}$ & 0.71 & $16.32_{-0.39}^{+0.27}$ & 0.65 \\ 
Pegasus III & $16.13_{-0.46}^{+0.28}$ & 0.78 & $16.20_{-0.42}^{+0.27}$ & 0.75 & $16.30_{-0.27}^{+0.16}$ & 0.71 \\ 
Tucana III & $16.13_{-0.14}^{+0.17}$ & 0.80 & $16.19_{-0.12}^{+0.16}$ & 0.78 & $16.30_{-0.20}^{+0.14}$ & 0.73 \\ 
Hercules & $16.11_{-0.24}^{+0.19}$ & 0.86 & $16.12_{-0.22}^{+0.20}$ & 0.86 & $16.18_{-0.33}^{+0.27}$ & 0.84 \\ 
Leo IV & $16.06_{-0.38}^{+0.28}$ & 0.78 & $16.10_{-0.37}^{+0.29}$ & 0.76 & $16.14_{-0.21}^{+0.19}$ & 0.74 \\ 
\hline
\hline
\end{tabular}
\label{tab:groupA_table_UFD_0.05}
\end{table*}

In this section, we summarize the results of $D$ factors for all the known dSphs that we studied.
First, we provide the $D$-factor median and 68\% confidence intervals in Tables~\ref{tab:groupA_table_classicals_0.5} and \ref{tab:groupA_table_classicals_0.05} for the classical dSphs and Tables~\ref{tab:groupA_table_UFD_0.5} and \ref{tab:groupA_table_classicals_0.05} for all the ultrafaint dSphs, organized in descending order of $D_{\rm dSph}$ values. Tables~\ref{tab:groupA_table_classicals_0.5} and \ref{tab:groupA_table_UFD_0.5} correspond to integration angles of  $\alpha_{\rm int}=0.5^{\circ}$, while Tables~\ref{tab:groupA_table_classicals_0.05} and \ref{tab:groupA_table_UFD_0.05} were obtained with $\alpha_{\rm int}=0.05^{\circ}$. Additionally, the relative contribution of the Milky-Way $D$ factor is shown for all the dSphs as a fractional contribution $R_{D_{\rm MW}} \equiv D_{\rm MW}/(D_{\rm MW}+D_{\rm dSph})$. In the case of the classical dSphs, the posterior densities were only calculated by employing $V_{\rm peak}>25$~km~s$^{-1}$, while a few values of $V_{\rm peak}$ are explored for the ultrafaint dSphs.

\begin{figure*}
    \centering
    \includegraphics[width=8cm]{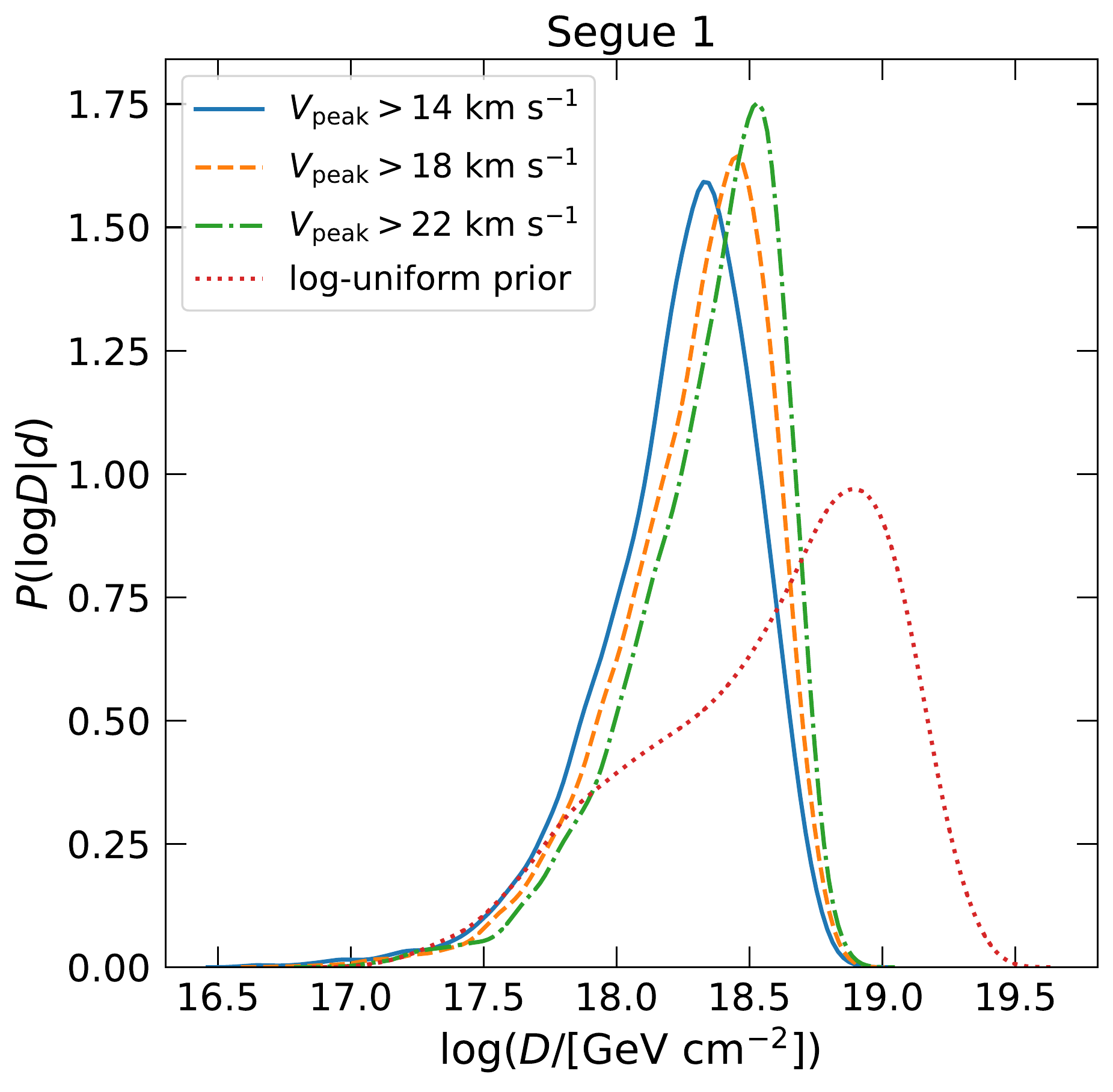}
    \includegraphics[width=8cm]{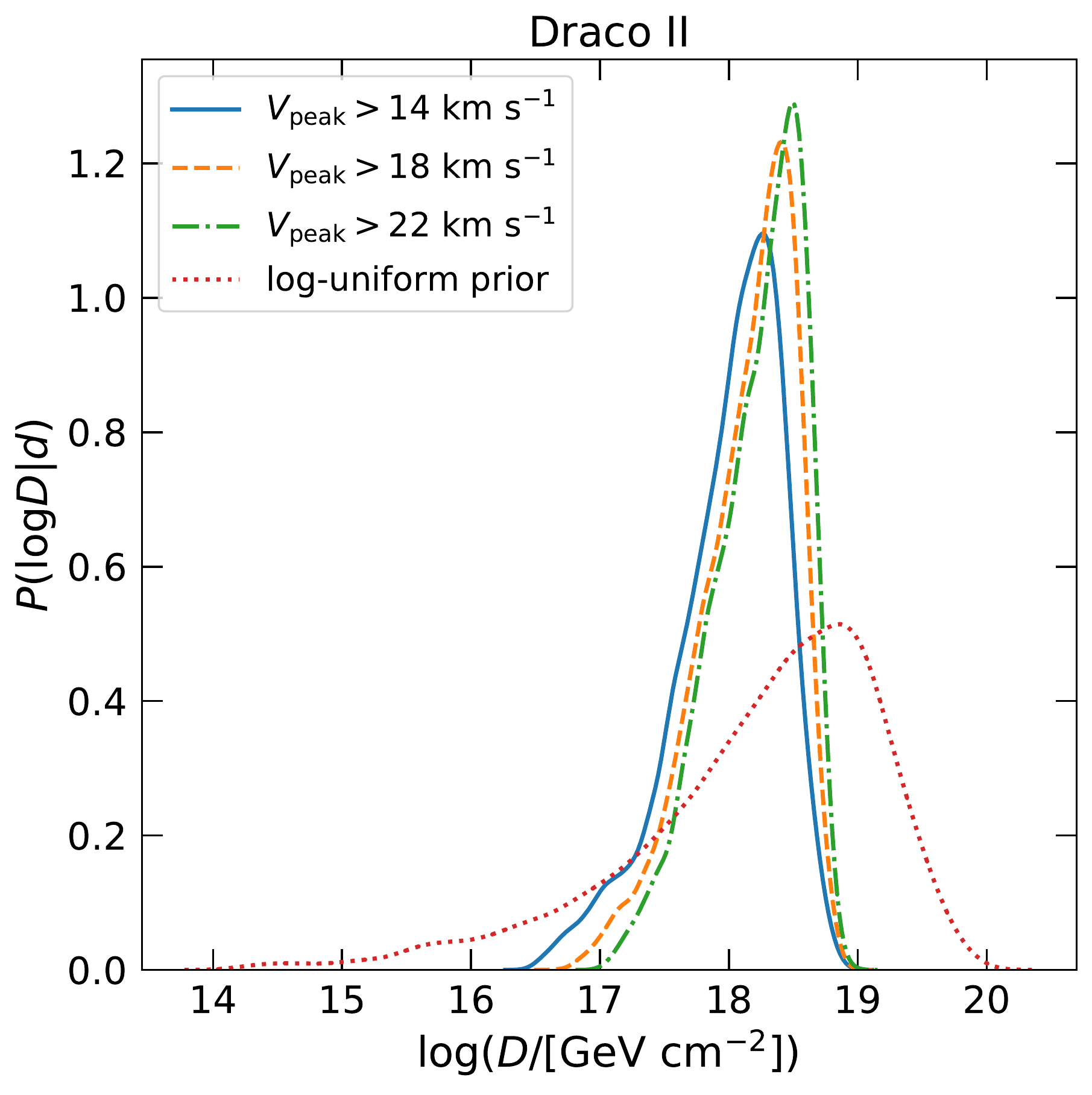}
    \includegraphics[width=8cm]{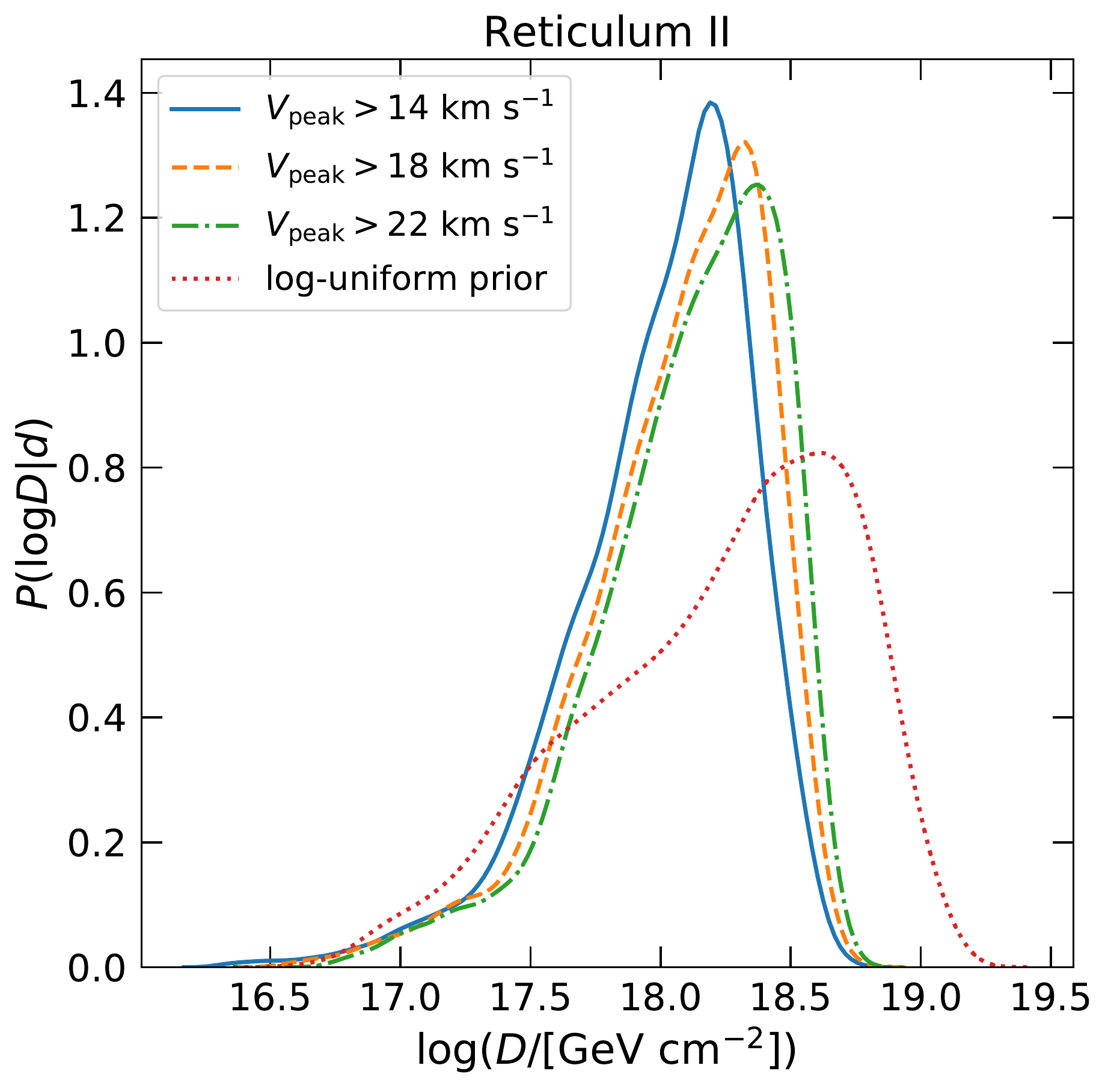}
    \includegraphics[width=8cm]{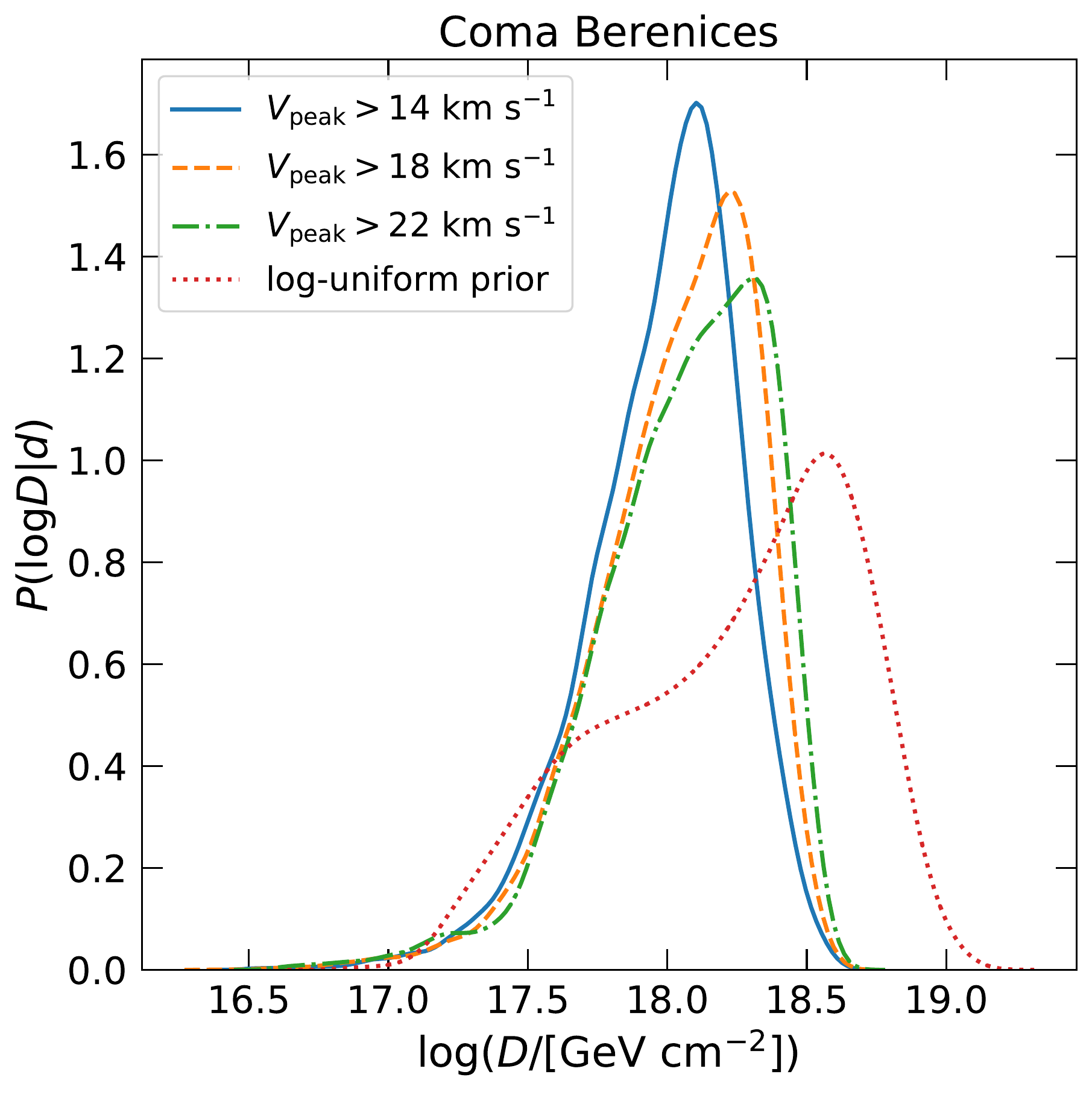}
    \caption{Posterior distributions of $D_{\rm dSph}(0.5^{\circ})$ for Segue 1, Draco II, Reticulum II, Coma Berenices, and Ursa Major I, obtained with satellite priors with $V_{\rm peak}>14~\rm km~s^{-1}$ (solid), $V_{\rm peak}>18~\rm km~s^{-1}$ (dashed), $V_{\rm peak}>22~\rm km~s^{-1}$ (dot-dashed), and log-uniform priors (dotted).}
    \label{fig:groupA_5_Dfactors}
\end{figure*}

Subsequently, in order to provide further visualization of the characteristics of the posterior distributions resulting from the parameter inference analysis, in Fig.~\ref{fig:groupA_5_Dfactors} we present the $D$-factor posterior distributions as shown in Fig.~\ref{fig:D_distribution_UMa_II} for some of the most promising dwarf galaxies for $\alpha_{\rm int}=0.5^{\circ}$.

\begin{figure*}
    \centering
    \includegraphics[width=8.5cm]{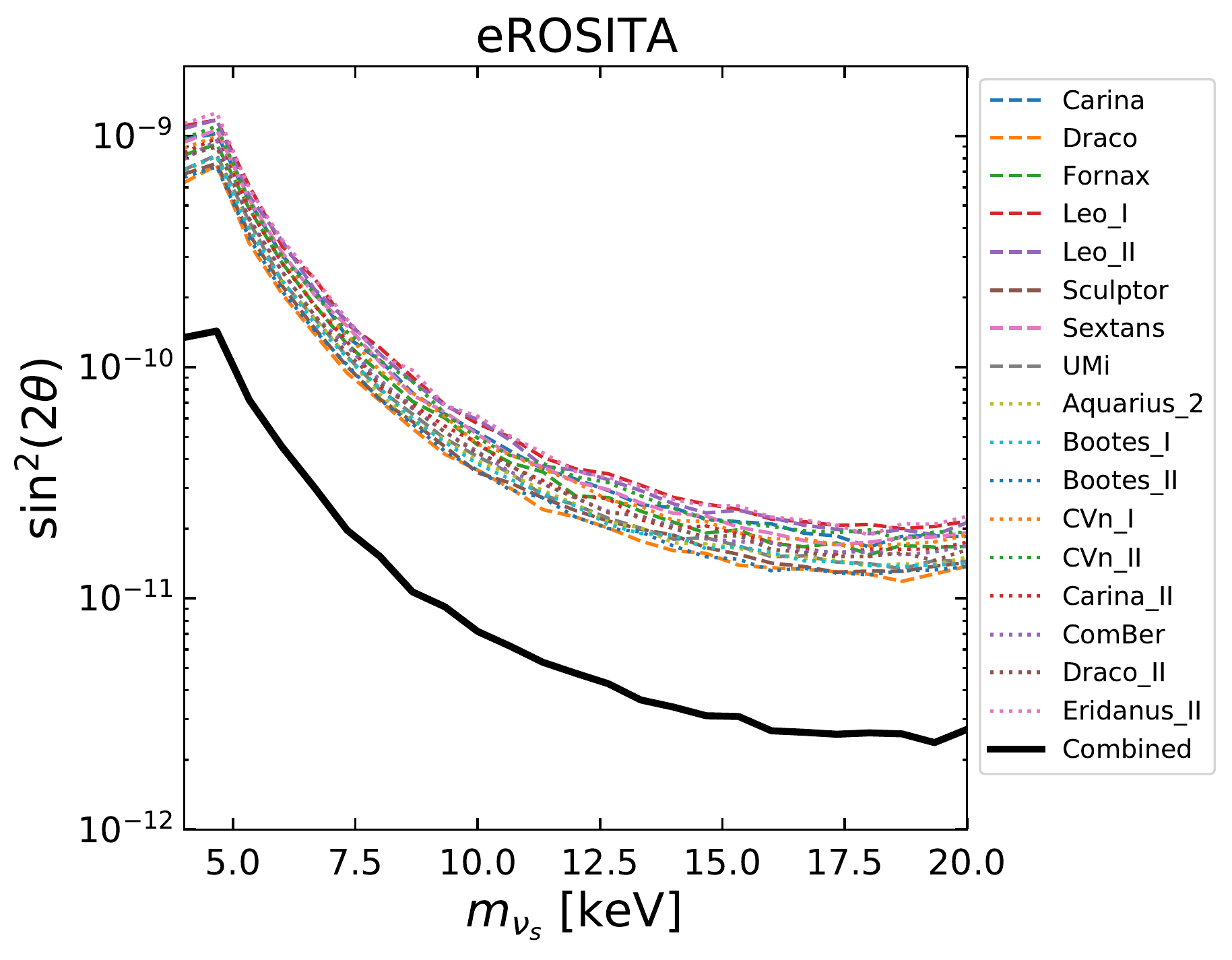}
    \includegraphics[width=8.5cm]{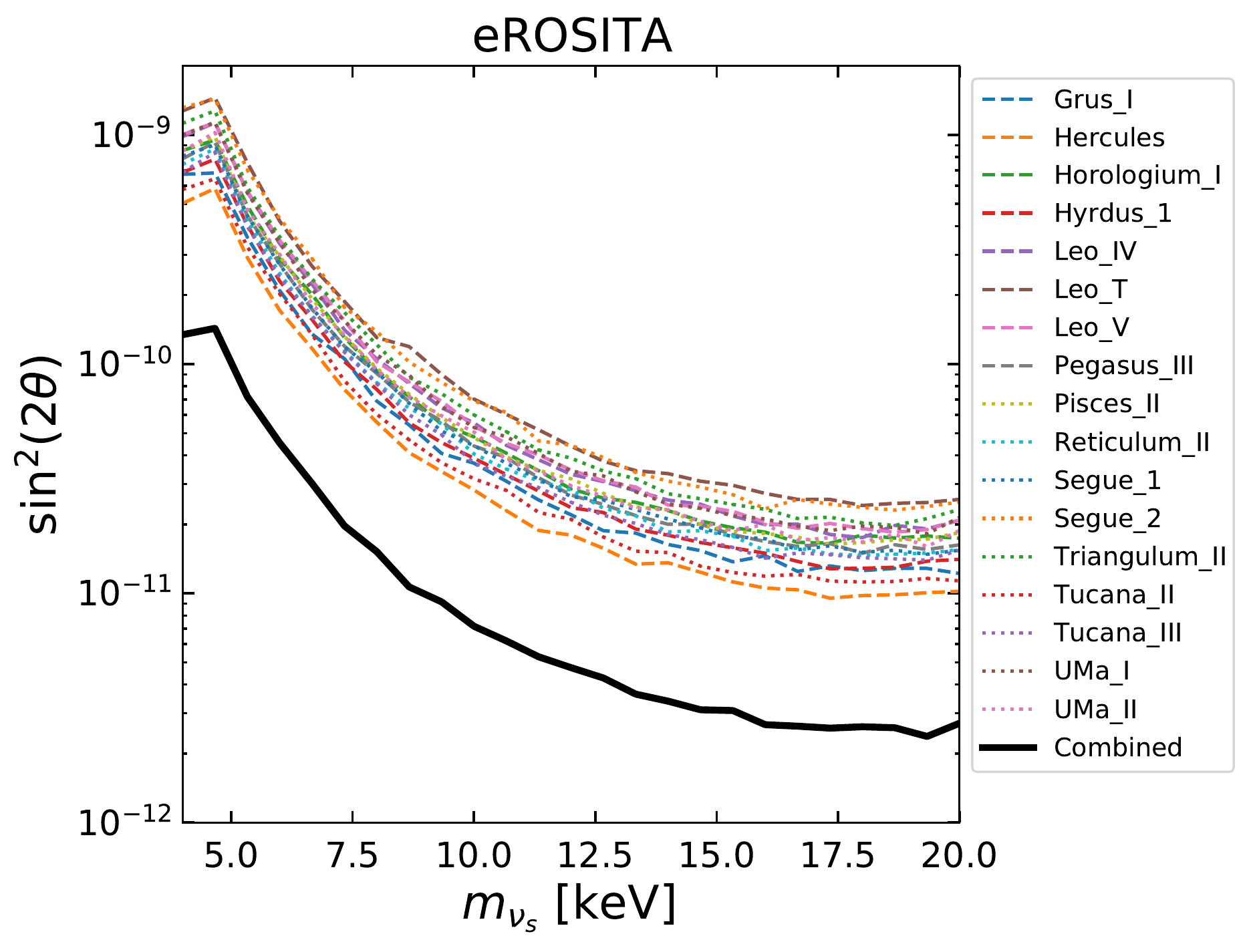}
    \caption{Medians of the mixing angle constraints with $\alpha_{\rm int} = 0.5^\circ$ for eROSITA computed for individual dSphs compared with the combined limits (thick solid).}
    \label{fig:groupA_eROSITA}
\end{figure*}

\begin{figure*}
    \centering
    \includegraphics[width=8cm]{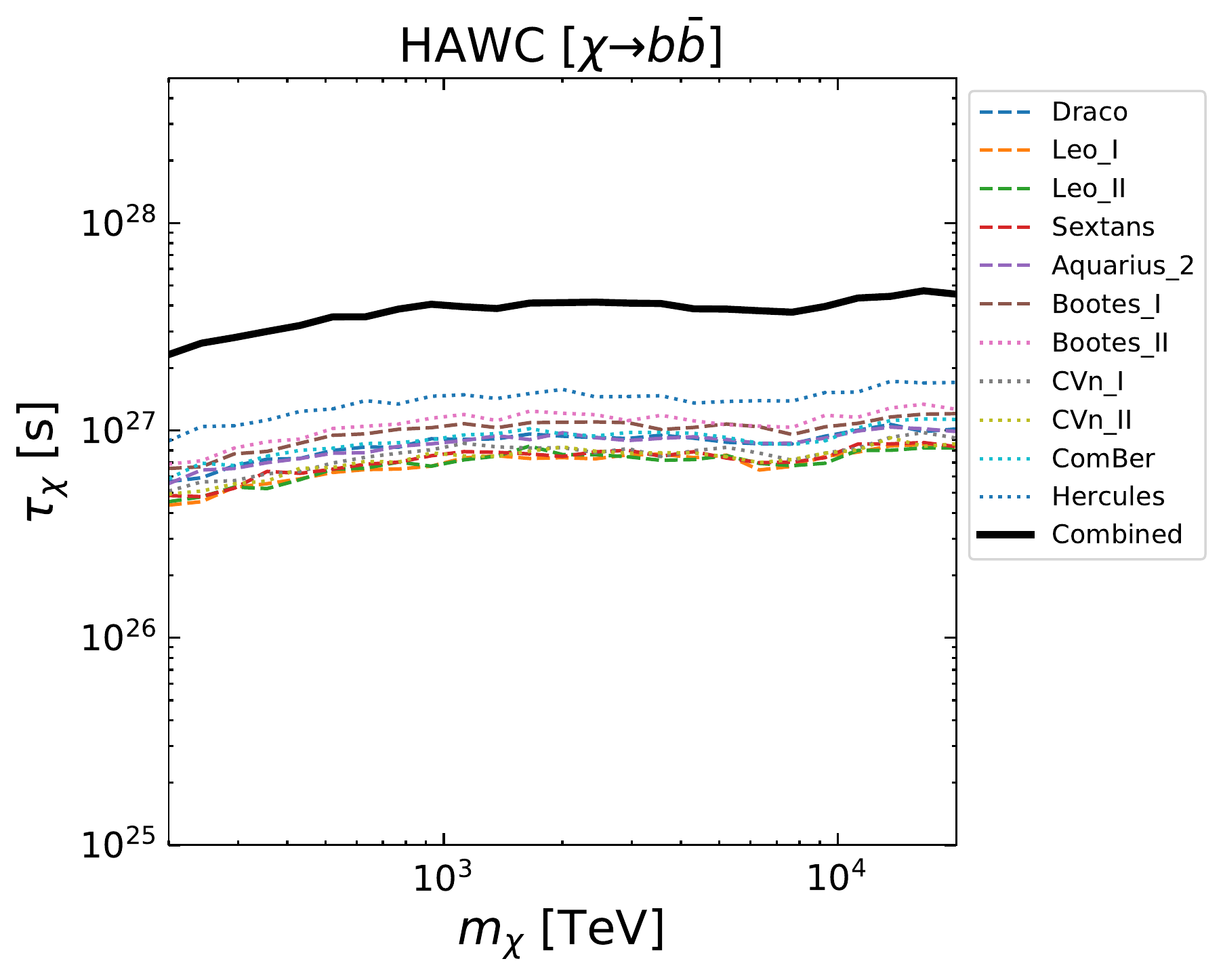}
    \includegraphics[width=8cm]{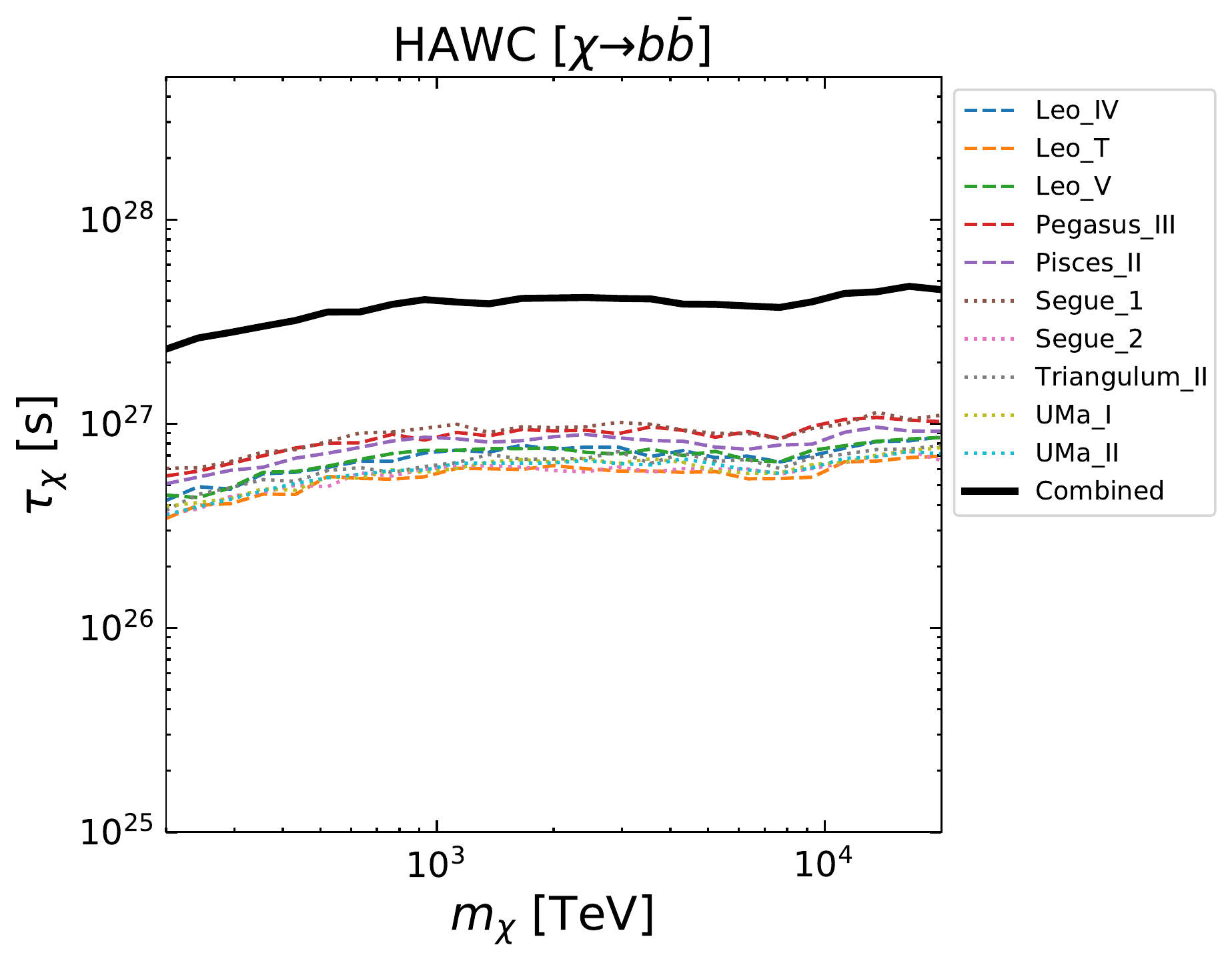}
    \includegraphics[width=8cm]{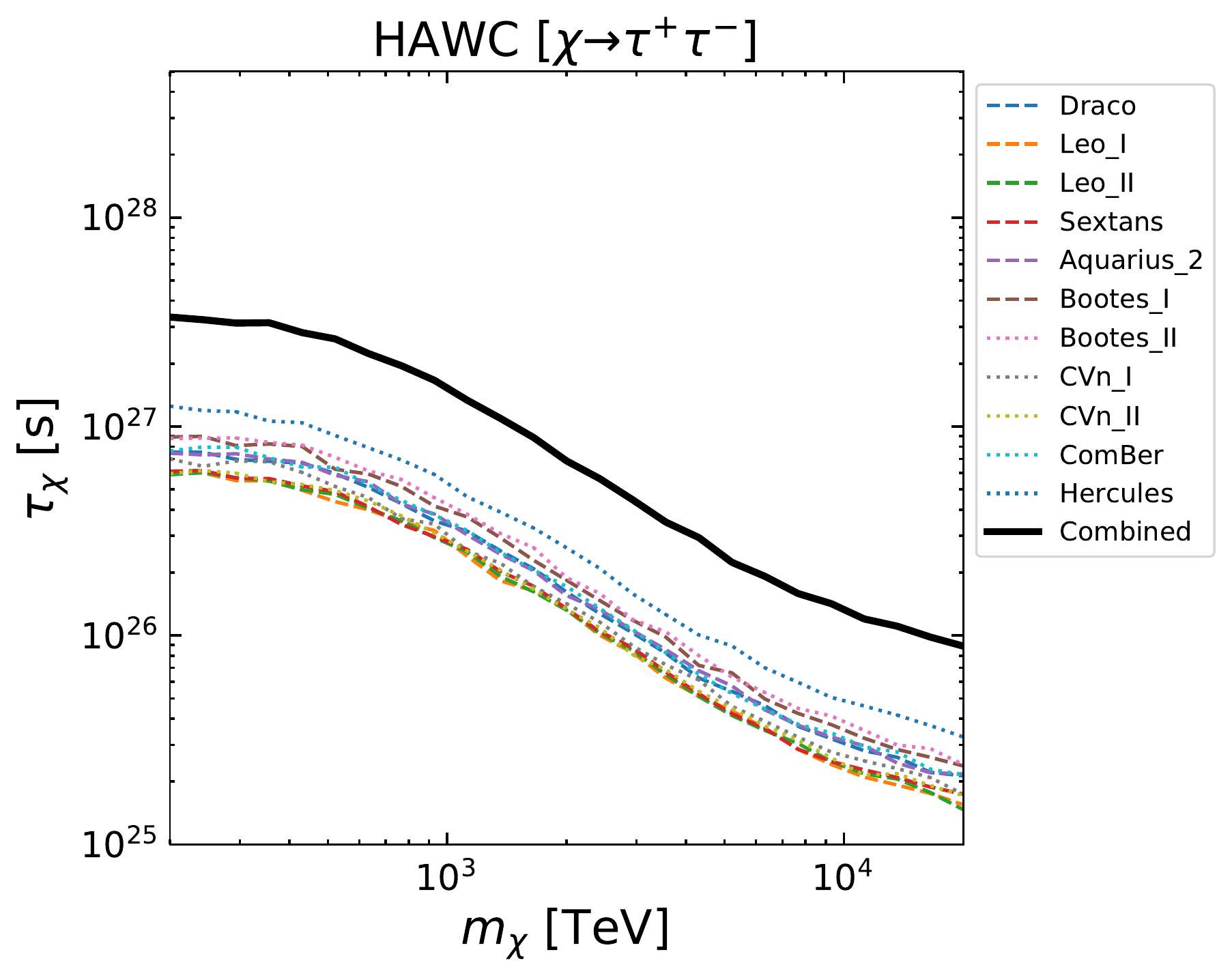}
    \includegraphics[width=8cm]{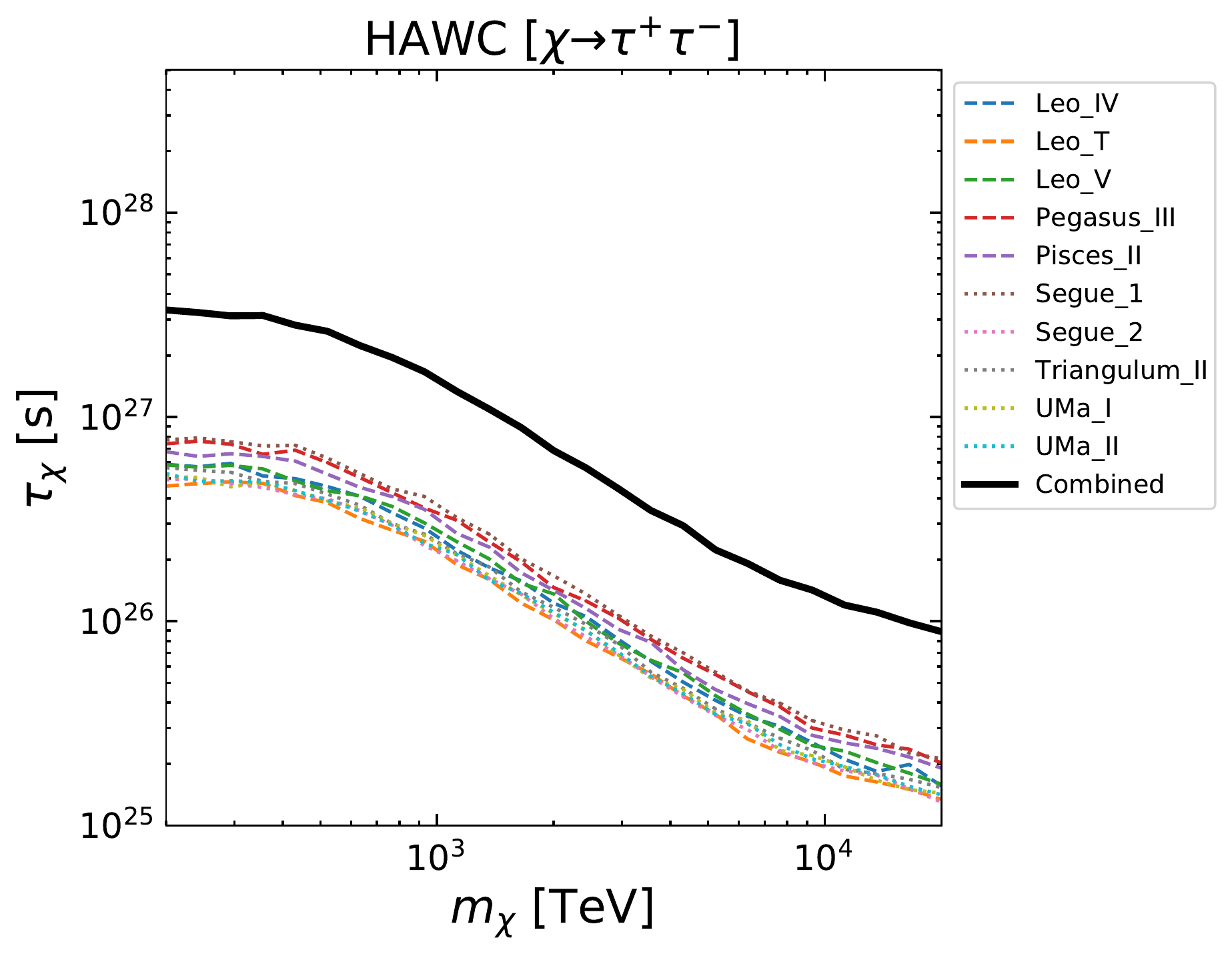}
    \caption{Medians of the dark matter lifetime constraints with $\alpha_{\rm int} = 0.5^\circ$ for HAWC computed for individual dSphs compared with the combined limits (thick solid). Top and bottom panels show the results of $b\bar{b}$ and $\tau^+\tau^-$ final states, respectively.}
    \label{fig:groupA_HAWC}
\end{figure*}

Lastly, we provide the constraints on the sterile neutrino mixing angle $\sin^22\theta$ for the eROSITA, and dark matter lifetime for the HAWC, resulting from the individual likelihood analysis of the known dSphs in Figs.~\ref{fig:groupA_eROSITA} and \ref{fig:groupA_HAWC}, respectively. 
In these figures, We show only the medians of the constraints.
These figures demonstrates that the overall constraints are much improved by gathering the statistical power of each individual dSph analyses.

\bibliography{bib.bib}

\end{document}